\documentclass[10pt,twocolumn,titlepage]{article}
\usepackage{geometry}
\usepackage[sectionbib,numbers,sort&compress]{natbib} 
\usepackage[version=3]{mhchem}
\usepackage{times}
\usepackage{balance} 
\usepackage{siunitx}
\usepackage{graphicx} 
\usepackage{lastpage}
\usepackage{fancyhdr}
\usepackage{color}

\newcommand{\kb}{$k_\mathrm{B}$}
\newcommand{\ids}{$I_\mathrm{ds}$}
\newcommand{\isc}{I_\mathrm{sc}}
\newcommand{\iphoto}{$I_\mathrm{photo}$}
\newcommand{\idark}{I_\mathrm{dark}}

\newcommand{\vds}{$V_\mathrm{ds}$}
\newcommand{\voc}{V_\mathrm{oc}}
\newcommand{\vg}{$V_\mathrm{g}$}

\newcommand{\mobility}{\si{\square\cm\per\volt\per\second}}
    \newcommand{\seebeck}{\si{\uV\per\kelvin}}
    \newcommand{\respo}{\si{\mA\per\W}}
    
    \newcommand{\tnumb}[1]{\tablenum[table-format = 1]{#1}}
    \newcommand{\tnumbb}[1]{\tablenum[table-format = +2]{#1}}
    \newcommand{\tnumbbb}[1]{\tablenum[retain-zero-exponent = false, table-format = +3.1e-1]{#1}}
    \newcommand{\tnumbbbb}[1]{\tablenum[retain-zero-exponent = false, table-format = +3.1e-1]{#1}}
    
    \newcommand{\tnrr}[1]{\tablenum[retain-zero-exponent = false, table-format = 1.2]{#1}}

\usepackage{booktabs}
\usepackage{pdflscape}
\usepackage{multirow}
\usepackage{floatrow}
\usepackage{amsfonts}
\usepackage{epstopdf}

\begin{document}
\thispagestyle{empty}
\twocolumn[

\begin{@twocolumnfalse}

\begin{center}

{\huge\textsc{Photocurrent generation with two-dimensional van der Waals semiconductors}}

\bigskip 

{\textbf{Michele Buscema,$^{\ast}$\textit{$^{a}$} Joshua O. Island,\textit{$^{a}$} Dirk J. Groenendijk,\textit{$^{a}$} Sofya I. Blanter,\textit{$^{a,b}$} Gary A. Steele\textit{$^{a}$}, Herre S.J. van der Zant\textit{$^{a}$}, and Andres Castellanos-Gomez$^{\ast}$\textit{$^{a,c}$}}}

\bigskip

{\textit{$^{a}$~ Kavli Institute of Nanoscience, Delft University of Technology, Lorentzweg 1, 2628 CJ, Delft, The Netherlands. \\ E-mail: m.buscema@tudelft.nl}}

{\textit{$^{b}$~ Current address: Institute for Experimental and Applied Physics, University of Regensburg, 93040 Regensburg, Germany.}}

{\textit{$^{c}$~  Instituto Madrile{\~n}o de Estudios Avanzados en Nanociencia (IMDEA-Nanociencia), 28049 Madrid, Spain.\\ E-mail: andres.castellanos@imdea.org}}

\bigskip

\Large{Abstract}
\end{center}

Two-dimensional (2D) materials have attracted a great deal of interest in recent years. This family of materials allows for the realization of versatile electronic devices and holds promise for next-generation (opto)electronics. Their electronic properties strongly depend on the number of layers, making them interesting from a fundamental standpoint. For electronic applications, semiconducting 2D materials benefit from sizable mobilities and large on/off ratios, due to the large modulation achievable via the gate field-effect. Moreover, being mechanically strong and flexible, these materials can withstand large strain ($>$ 10\%) before rupture, making them interesting for strain engineering and flexible devices. Even in their single layer form, semiconducting 2D materials have demonstrated efficient light absorption, enabling large responsivity in photodetectors. Therefore, semiconducting layered 2D materials are strong candidates for optoelectronic applications, especially for photodetection. Here, we review the state-of-the-art in photodetectors based on semiconducting 2D materials, focusing on the transition metal dichalcogenides, novel van der Waals materials, black phosphorus, and heterostructures.
\end{@twocolumnfalse}
\vspace{1 cm}
]

\tableofcontents


\vspace{0.5cm}
\section{Introduction}

Photodetectors are a key component of many devices we use in our daily life. From imaging to optical communications, we rely on photodetectors to convert the information stored in light into electrical signals that can be processed by standard electronics. Silicon photodetectors are readily integrated in complementary metal-oxide-semiconductor (CMOS) technology, profiting from device miniaturization and scalability which considerably reduce costs and expand the range of applications.

Bulk Silicon photodetectors suffer from the limitations of silicon as a light-absorbing material. Its indirect bandgap of about \SI{1.1}{\eV} limits absorption to the visible and near-infrared part of the electromagnetic spectrum and reduces its efficiency. To achieve sizable responsivities, photodetectors based on bulk silicon rely on a thick channel, making the photodetector fully opaque. Silicon is also a brittle material in bulk, precluding its use in bendable device concepts. Other 3D materials, like InGaAs and related heterostructures, are also currently employed in photodetectors. While they allow efficient detection of IR wavelengths, they share most of silicon drawbacks and add a sizable increase in cost due to fabrication complexity. 

Novel, nanoscale semiconducting materials are needed to overcome the limitation of bulk silicon for photodetection. For instance, nanostructured silicon\cite{Menon2015} and especially Si nanowires\cite{Mulazimoglu2013} have shown in the past years the possibility to realize bendable and transparent devices. More recently, silicene (a single layer of silicon atoms) has been successfully used to fabricate FETs, opening the door to the study its opto-electronic properties.\cite{Tao2015}

These new materials should provide higher absorption efficiencies in the visible, an extended operation wavelength range, fast detection and, preferably, facile integration with current CMOS technology for readout. 

The large class of semiconducting layered materials presents appealing properties such as high transparency (and yet strong light-matter interaction), flexibility and ease of processing. Therefore, they recently attracted a large research effort aimed at understanding the principal photodetection mechanisms and device performances.

The isolation of graphene in 2004,\cite{Novoselov2004,Novoselov2005a} rapidly followed by the discovery of its amazing properties,\cite{Novoselov2005, Novoselov2006, Novoselov2007, Lee2008, Xia2009,Mueller2010, Wang2012,Wang2013} has generated an intense research effort on layered 2D materials.\cite{Novoselov2005,Coleman2011} In layered 2D materials, the atoms forming the compound are arranged into planes (layers) that are held together by strong in-plane bonds, usually covalent. To form a 3D crystal, the atomic layers are stacked in the out-of-plane direction with weak van der Waals interactions. This allows the exfoliation of bulk crystals and the fabrication of thinner \emph{flakes}, even down to the single layer limit. \cite{Novoselov2005a} 

Flakes of layered materials present several advantages over conventional 3D materials for photodetection, from both a practical and a fundamental standpoint. The atomic thickness renders these materials almost transparent, which is of high interest for novel applications, e.g. photovoltaics integrated in fa\c{c}ades or wearable electronics. Their atomic thickness is also responsible for quantum confinement effects in the out-of-plane direction. These effects are particularly strong in the semiconducting transition metal dichalcogenides - TMDCs -  where the reduced thickness results in strongly bound excitons, increasing the absorption efficiency.\cite{Mak2010,Mak2012,Mak2013} Another effect of the vertical confinement is the modulation of the bandgap as a function of the number of layers, particularly evident in the TMDCs,\cite{Ayari2007,Radisavljevic2011,Mak2010,Splendiani2010,Eda2011,Yun2012} which modulates the optical absorption edge.

In 2D semiconductor crystals, the electronic bands are localized in nature (in the case of the TMDCS due to the $d$  orbital contribution), leading to sharp peaks in the density of states at a particular energy (i.e. Van Hove singularities)\cite{Britnell2013}. In several 2D semiconductors (for instance \ce{MoS2}, \ce{WS2} and \ce{WSe2}) these singularities happen close to the conduction and valence band edges. Therefore, a photon with energy close to the bandgap has an increased probability to excite an electron-hole pair due to the large availability of empty states given by the diverging DOS at the singularity. Thus, despite of the reduced thickness of 2D semiconductors (yielding the high transparency) they strongly interact with the incident light.

2D materials have also demonstrated remarkable elastic modulus and large strain ($> 10 \% $) before rupture. Large strains have a strong effect on the electronic and optical behavior of these materials.\cite{Lee2008,Castellanos-Gomez2012,Yun2012,He2013,Conley2013,Zhu2013} Thus, strain-engineering can be used to tune the optical properties and realize novel device architectures (weareable, bendable devices) and devices with novel functionalities, like exciton funnelling. \cite{Feng2012, Castellanos-Gomez2013d}

In this Review, we present the state-of-art in photodetection with layered materials, especially focusing on the semiconducting 2D materials. For the purpose of this review, we consider as two dimensional the materials in which (i) the atoms have strong in-plane bonds, (ii) the atoms are arranged into planes with high crystalline order and (iii) these atomically-thin planes are held together only by van der Waals interactions in the vertical direction (i.e. there are no dangling bonds at the surfaces of the planes). 
The absence of dangling bonds between the planes and the weak van der Waals interaction makes it possible to isolate single-unit-cell-thickness flakes by mechanical exfoliation of a parent, bulk crystal. They also allow stacking several layers of different materials on top of each other and to grow them on a variety of substrates, while still preserving the high crystalline order. These are all properties in stark contrast with conventional 3D crystalline materials.

Layered materials, even when relatively thick (\SI{\sim 10}{\nm}) compared to a single layer, already possess a strong in-plane vs out-of-plane anisotropy in their properties. Further thickness reduction to a small number of layers, ultimately to a single layer, also induces quantum confinement in the vertical direction, which has a strong impact on the bandgap of the material. This effect is particularly strong in the TMDCs, whose bandgap increases and becomes direct in single layer, and black phosphorus, whose bandgap increases of more than 3 times. 

In the following, we review devices based on flakes with thickness from one unit cell to several nanometers and with in-plane dimensions in the order of microns. We start by briefly introducing the main photodetection mechanisms and figures-of-merit that are useful to compare different photodetectors. We then move on to describe photodetectors based on semiconducting transition metal dichalcogenides (TMDCs). Next, we summarize the performance of photodetectors based on the novel van der Waals materials, such as \ce{Ga}, \ce{In} and \ce{Sn} chalcogenides and \ce{Ti}, \ce{Hf} and \ce{Zr} trichalcogenides. Then we describe the recent progress in black phosphorus-based photodetectors. We then briefly compare the responsivity and time response of the reviewed detectors. Subsequently, we describe a few of the future directions of photodetection with layered materials. Finally, we summarize the main conclusions of this Review and list future challenges.

\section{Photocurrent generation mechanisms}\label{theory}

This Section briefly introduces the mechanisms enabling photodetection. We start by describing the mechanisms driven by electric field separation of electron-hole pairs generated by photon absorption. These mechanisms are usually categorized as photoconduction, photogating and the photovoltaic effect and are especially relevant for (photo)-field-effect transistors (FETs). Next, we introduce thermal processes (photo-thermoelectric, bolometric) that can also generate or modulate the photoresponse in photodetectors. At last, we discuss the relevant figures-of-merit for photodetectors to facilitate comparison among devices working with different principles.

\subsection{Photocurrent generation driven by electron-hole separation}

\subsubsection{Photoconductive effect.~~} 
\begin{figure}[h!]
\captionsetup{width=.9\textwidth}
\centering
\includegraphics[width=\textwidth]{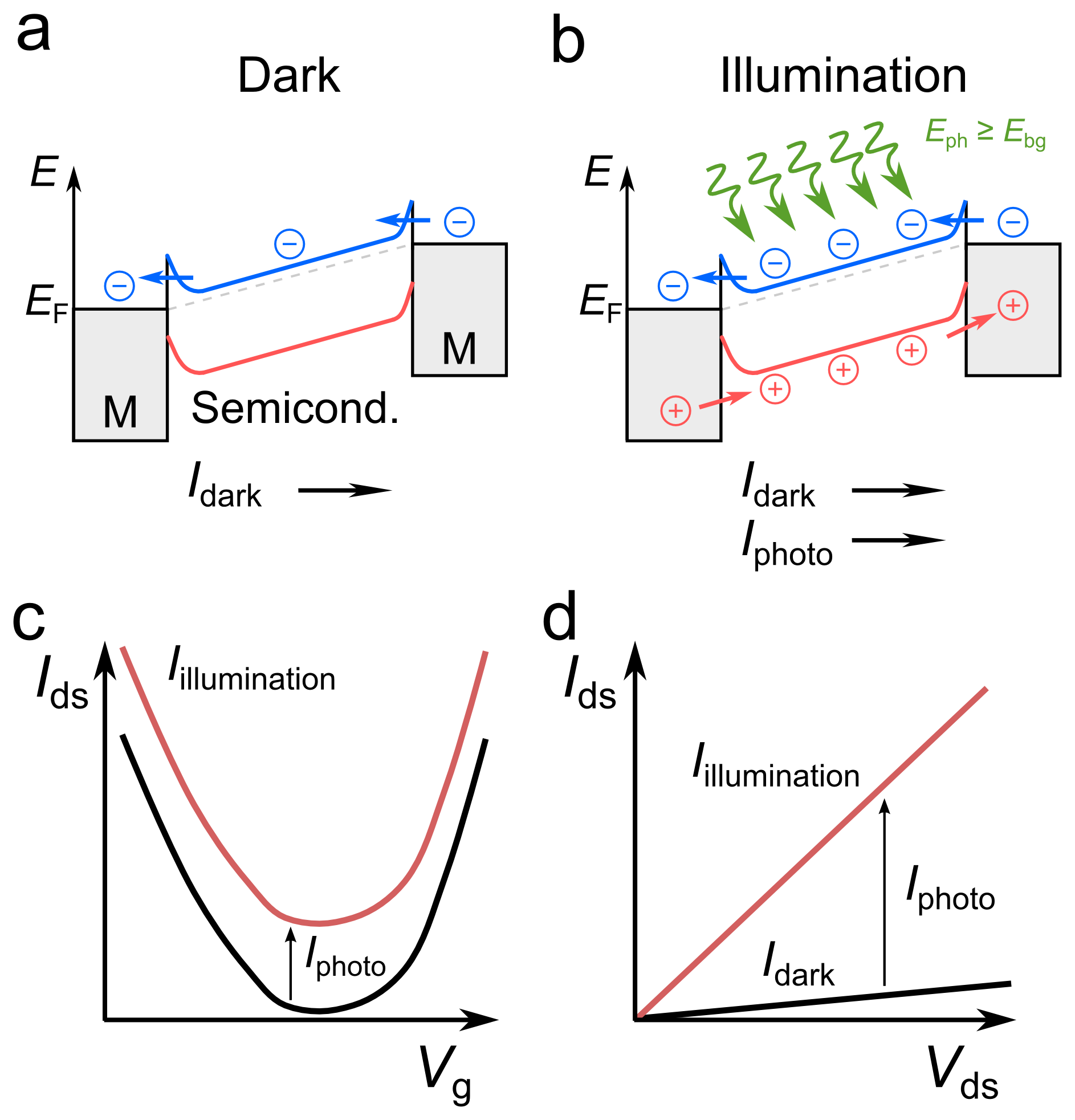}
\caption{Schematic of the photoconductive effect. (a) Band alignment for a semiconductor channel contacted with two metals (M) under an external bias without illumination. A small current flows through the device ($I_\mathrm{dark}$). (b) Band alignment under illumination with photons of energy ($E_\mathrm{ph}$) higher  than the bandgap ($E_\mathrm{bg}$). The absorption of photons generates e-h pairs that are separated by the external applied bias, generating a photocurrent ($I_\mathrm{photo}$) which adds to $I_\mathrm{dark}$. (c) \ids$-$\vg\ traces in the dark (solid black line) and under illumination (solid red line). Illumination results in an increase in the conductivity (vertical shift) and a positive photocurrent across the entire gate voltage range. (d) \ids$-$\vds\ curves in the dark (solid black line) and under illumination (solid red line). Illumination results in an increase in the conductivity and a positive photocurrent.}
\label{intro_PC}
\end{figure}

In the photoconductive effect, photon absorption generates extra free carriers, reducing the electrical resistance of the semiconductor (see Figure \ref{intro_PC}).\cite{Rose1963,Bube1992,Saleh1991,Konstantatos2010} Figure \ref{intro_PC}a sketches the band diagram of an FET. Without illumination and under an applied bias (\vds), a small source-drain current can flow ($\idark$). Under illumination, the absorption of photons with energy higher than the bandgap ($E_\mathrm{ph} > E_\mathrm{bg}$) generates e-h pairs which are separated by the applied \vds\ (Figure \ref{intro_PC}b). The photogenerated free electrons and holes drift in opposite directions towards the metal leads, resulting in a net increase in the current (\iphoto). This photogenerated current adds to the dark current, reducing the resistance of the device, as depicted in Figure \ref{intro_PC}c and Figure \ref{intro_PC}d.

It is instructive to consider the case of a large difference between the electron and hole mobilities, resulting in a large difference in the electron/hole transit time ($\tau_\mathrm{transit}$):\cite{Saleh1991}
\begin{equation}\label{transit}
\tau_\mathrm{transit} = \frac{L^2}{\mu \cdot V_\mathrm{ds}},
\end{equation}
where $L$ is the length of the transistor channel, $\mu$ the charge carrier mobility and $V_\mathrm{ds}$ the source-drain bias. If the hole mobility is much lower than the electron mobility, the photogenerated  electrons can cross the channel much faster than the photogenerated holes. Until recombination or hole extraction, many electrons can participate in the photocurrent, leading to the photoconductive gain ($G$). Effectively, this means that more electrons can be extracted from a single photon, resulting in a quantum efficiency larger than one. The photoconductive gain is the ratio of the photogenerated carrier lifetime ($\tau_\mathrm{photocarriers}$) and the transit time:\cite{Saleh1991}
\begin{equation}\label{gain}
G = \frac{\tau_\mathrm{photocarriers}}{\tau_\mathrm{transit}} = \frac{\tau_\mathrm{photocarriers} \cdot\mu \cdot V}{L^2}.
\end{equation}

Hence, large $\tau_\mathrm{photocarriers}$ and a large mismatch in the electron/hole mobility yield large $G$.

\begin{figure}[h!]
\captionsetup{width=.9\textwidth}
\centering
\includegraphics[width=\textwidth]{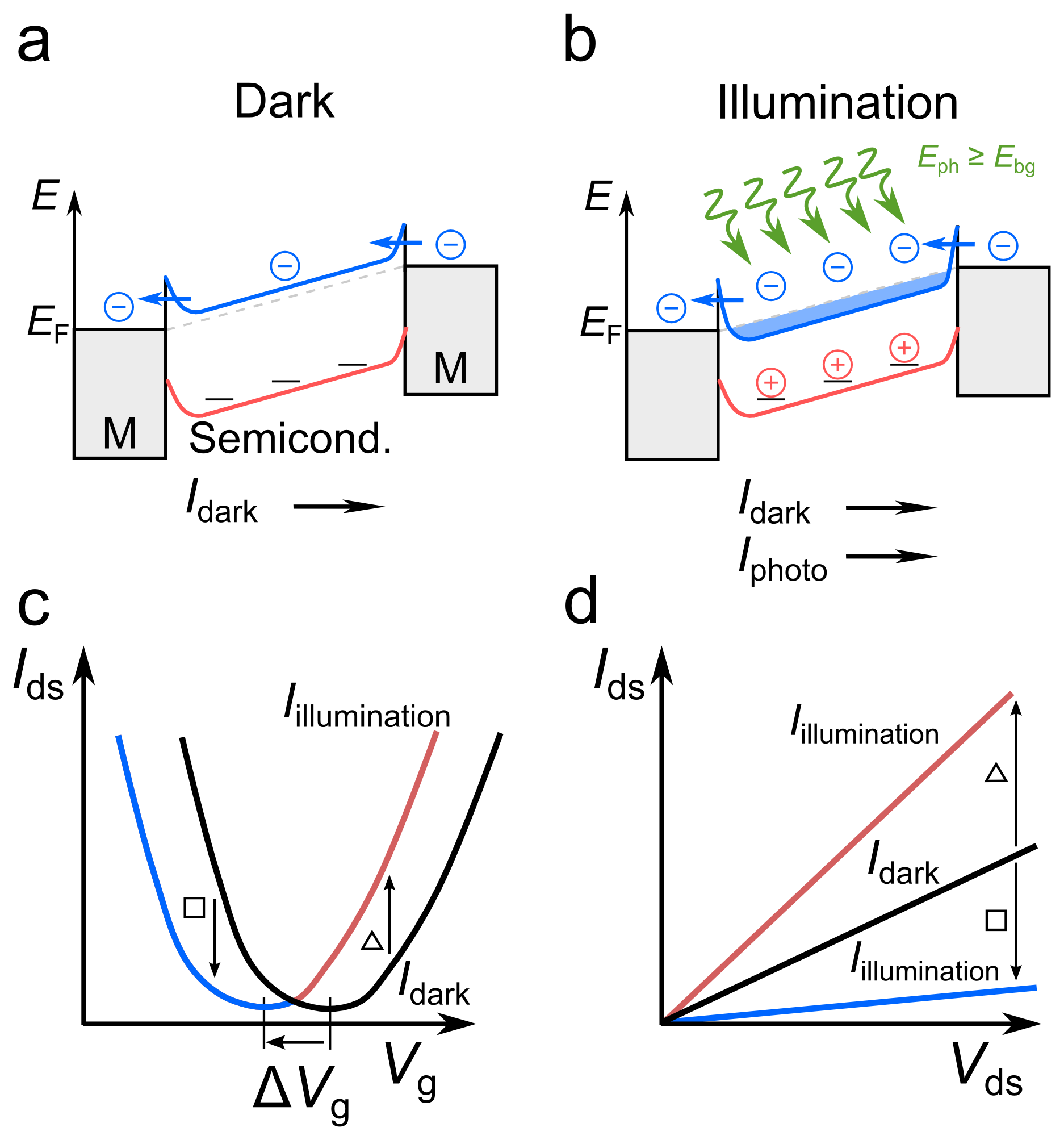}
\caption{Photogating schematics. (a) Band alignment for a semiconductor channel contacted with two metals (M) under external bias in the dark. A small electron current flows through the device ($I_\mathrm{dark}$). The solid horizontal solid segments represent trap states at the valence band edge. (b) Band alignment under illumination with photons of energy ($E_\mathrm{ph}$) higher than the bandgap ($E_\mathrm{bg}$). The absorption of a photon generates an electron-hole pair. Holes are trapped at the band edge and act as a local gate. The field-effect induces more electrons in the channel, generating a photocurrent ($I_\mathrm{photo}$) which adds to the $I_\mathrm{dark}$. (c) \ids$-$\vg\ traces in the dark (solid black line) and under illumination (solid blue/red line). Illumination results in a horizontal shift ($\Delta V_\mathrm{g}$) of the traces, due to the trapped charges that act as an additional gate electric field. (d) \ids$-$\vds\ curves in dark (solid black line) and under illumination at the gate value marked by a triangle in panel c (solid red line) and at the gate voltage marked by a square in panel c (solid blue line). Illumination results in positive photocurrent for one gate value (triangle marker) and negative photocurrent for another gate value (square marker).}
\label{intro_Pgate}
\end{figure}

\emph{Photogating} is a particular example of the photoconductive effect. If holes/electrons are \emph{trapped} in localized states (see Figure \ref{intro_Pgate}a), they act as a local gate, effectively modulating the resistance of the material. In this case, $\tau_\mathrm{photocarriers}$ is only limited by the recombination lifetime of the localized \emph{trap} states, leading to a large $G$.\cite{Saleh1991}
The trap states where carriers can reside for long times are usually located at defects or at the surface of the semiconducting material. This effect is of particular importance for nanostructured materials, like colloidal quantum dots, nanowires and two dimensional semiconductors, where the large surface and reduced screening play a major role in the electrical properties.

Photogating can be seen as a horizontal shift in the \ids$-$\vg\ traces under illumination, as shown schematically in Figure \ref{intro_Pgate}. Under illumination, the absorption of a photon generates an e-h pair. One carrier type (holes in Figure \ref{intro_Pgate}b) is then trapped in localized states with energy near the valence band edge. In this case, the electric field-effect of the trapped holes shifts the Fermi level which, in turn, induces more electrons. The increased electron density reduces the resistance of the device, allowing more current to flow (\iphoto). Under illumination, the \ids$-$\vg\ trace will be horizontally shifted ($\Delta V_\mathrm{g}$) with respect to the dark trace due to the effective gate electric field of the trapped charges. The sign of $\Delta V_\mathrm{g}$ indicates the polarity of the trapped carrier. Photogating can also result in a negative \iphoto, as sketched in Figure \ref{intro_Pgate}c,d. 

In practice, the clear distinction between photoconductive and photogating effect is faded, since both effects can take place in the same device. However, the difference in their time scales can be used to disentangle their signatures, as shown by Furchi \emph{et al.}.\cite{Furchi2014}

\subsubsection{Photovoltaic effect.~~} 
\begin{figure}[h!]
\captionsetup{width=.9\textwidth}
\centering\includegraphics[]{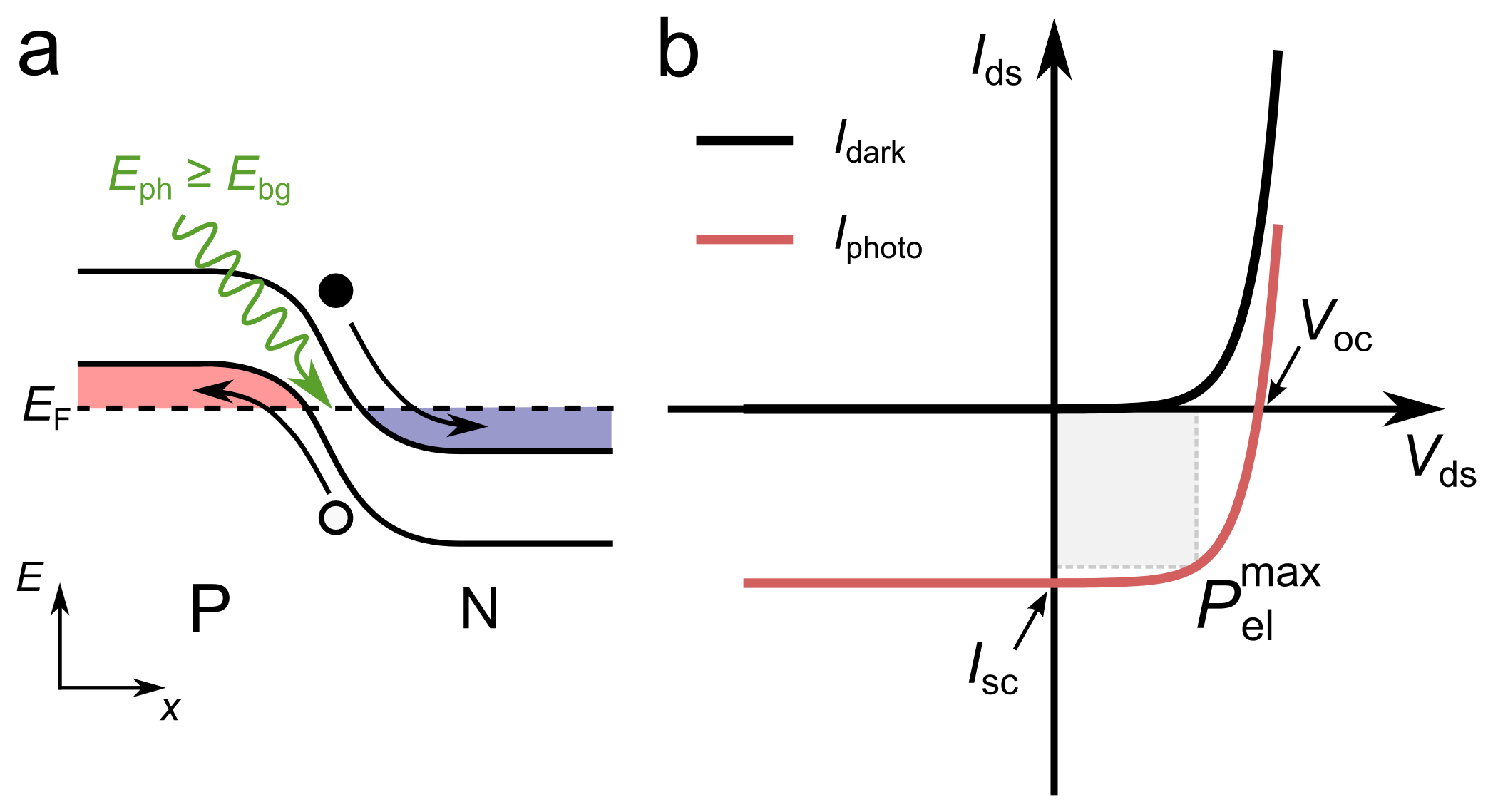}
\caption{Photovoltaic effect schematics. (a) Band alignment in a PN junction. The dashed line represents the common Fermi energy ($E_\mathrm{F}$) for the P-doped and N-doped semiconductor. The absorption of a photon with $E_\mathrm{ph} > E_\mathrm{bg}$ generates an electron-hole (black dot and white dot respectively) pair. The electron-hole pair is then separated and accelerated by the built-in electric field at the junction. (b)  \ids$-$\vds\ curves in the dark (solid black line) and under illumination (solid red line). The illumination results in a short-circuit current $\isc$ and an open-circuit voltage $\voc$. The device produces electrical power when operated in the fourth quadrant. The point of maximum power generation is indicated as $P_\mathrm{el}^\mathrm{max}$ }
\label{intro_PV}
\end{figure}

In the photovoltaic effect, photogenerated e-h pairs are separated by an \emph{internal} electric field. The origin of the internal electric field could be a PN junction or a Schottky barrier at the interface between a semiconductor and a metal (see Figure \ref{intro_PV}a).\cite{Saleh1991} In both cases, the devices present nonlinear \ids$-$\vds\ characteristics in the dark. In the case of PN junctions, the forward source-drain current \ids\ is exponential with the source-drain voltage \vds\ as $I_\mathrm{ds} \propto \exp(V_\mathrm{ds}) -1 $; the reverse current, on the other hand, is negligibly small until junction breakdown.\cite{Sze2006} Under illumination and zero external bias ($V_\mathrm{ds} = 0$), the internal electric field separates the photoexcited e-h pairs thereby generating a sizable photocurrent (short-circuit current, $\isc$). Keeping the circuit open leads to the accumulation of carriers of opposite polarities in distinct parts of the device, thereby generating a voltage (open circuit voltage, $\voc$). Under illumination and reverse bias, the magnitude of the reverse current increases since the photoexcited carriers are swept in opposite directions by the junction electric field. An additional forward-bias is required to compensate the photogenerated reverse-current, also justifying a non-zero $\voc$. In contrast to the photoconductive effect, the photovoltaic effect can also be used to convert the energy of the photons to electrical energy.

Figure \ref{intro_PV}b plots the \ids$-$\vds\ characteristics of a PN junction in the dark (black solid line) and under illumination (red line). Since the photocurrent has the same sign as the reverse current, the \ids$-$\vds\ under illumination appears shifted downwards, with respect to the dark curve. The part of the \ids$-$\vds\ curve that lays in the fourth quadrant (negative current, positive bias) is used to generate electrical power ($P_\mathrm{el}^\mathrm{max}$).

Photodetectors based on the photovoltaic effect are usually PN diodes and are used at zero bias (photovoltaic mode) or under reverse bias (photoconductive mode). In the photovoltaic mode, the dark current is the lowest, improving the detectivity (see Section \ref{PD_fom}) of the detector. However, the absolute responsivity (see Section \ref{PD_fom}) is usually smaller than a photodetector working with the photoconducting or photogating mechanism, since there is no internal gain. The advantage of reverse-bias operation (the photoconductive mode) is the reduction of the junction capacitance, increasing the speed of the photodiode. Under large reverse bias, the strong junction electric field can give enough energy to the photogenerated electrons to initiate impact ionization multiplications, or avalanching (avalanche photodiode - APD). This mechanism provide large internal gain, so that light of extremely low power can be detected.

\subsubsection{Simplified photocurrent generation phenomenological description.~~}
The generated \emph{photocurrent, \iphoto,} in a photoconductive detector can be estimated with the following phenomenological formula:
\begin{equation}\label{photogeneration}
I_\mathrm{photo} = I_\mathrm{illumination} - I_\mathrm{dark}  = \Gamma \cdot \eta \cdot e \cdot G, 
\end{equation}
where $\Gamma$ is the number of absorbed photons per unit time, $\eta$ the efficiency of the conversion of the absorbed photons to electrons, $e$ the electron charge and $G$ is the photoconductive gain. The parameter $\eta$ is the internal quantum efficiency of the detector, without considering gain mechanisms. For a photovoltaic detector $G = 0$.

With the help of Equation \ref{gain}, we can now rewrite Equation \ref{photogeneration} as:\cite{Koppens2014}
\begin{equation}\label{photocurrent}
I_\mathrm{photo} = \Gamma \cdot \eta \cdot e \cdot G \frac{\tau_\mathrm{photocarriers} \cdot\mu \cdot V}{L^2}.
\end{equation}
From this Equation we see that \iphoto\ is linearly dependent on the photon flux (i.e. the excitation power), the photogenerated carrier lifetime, the electron mobility and the applied bias, following simple physical intuition.

This simplified model, however, does not take into account the presence of a finite number of trap states. These trap states strongly affect the dependence of the photocurrent on the excitation power (usually it becomes sub-linear) and the carrier lifetime (usually it increases). Engineering trap states is a viable way to achieve ultra high gain, usually at the expense of a slower time response of the detector.

\subsection{Photocurrent generation driven by thermal mechanisms}\label{sec_PV_PTE}
Temperature \emph{gradients} induced by non-uniform heating under illumination can also generate a photocurrent or photovoltage through the photo-thermoelectric effect. On the other hand, a \emph{homogeneous} temperature change affects the resistivity of a material (photo-bolometric effect), which can be detected by electrical means.

\subsubsection{Photo-thermoelectric effect.~~}\label{par_PTE} 
\begin{figure}[h!]
\captionsetup{width=.9\textwidth}
\centering
\includegraphics[width = 0.9\textwidth]{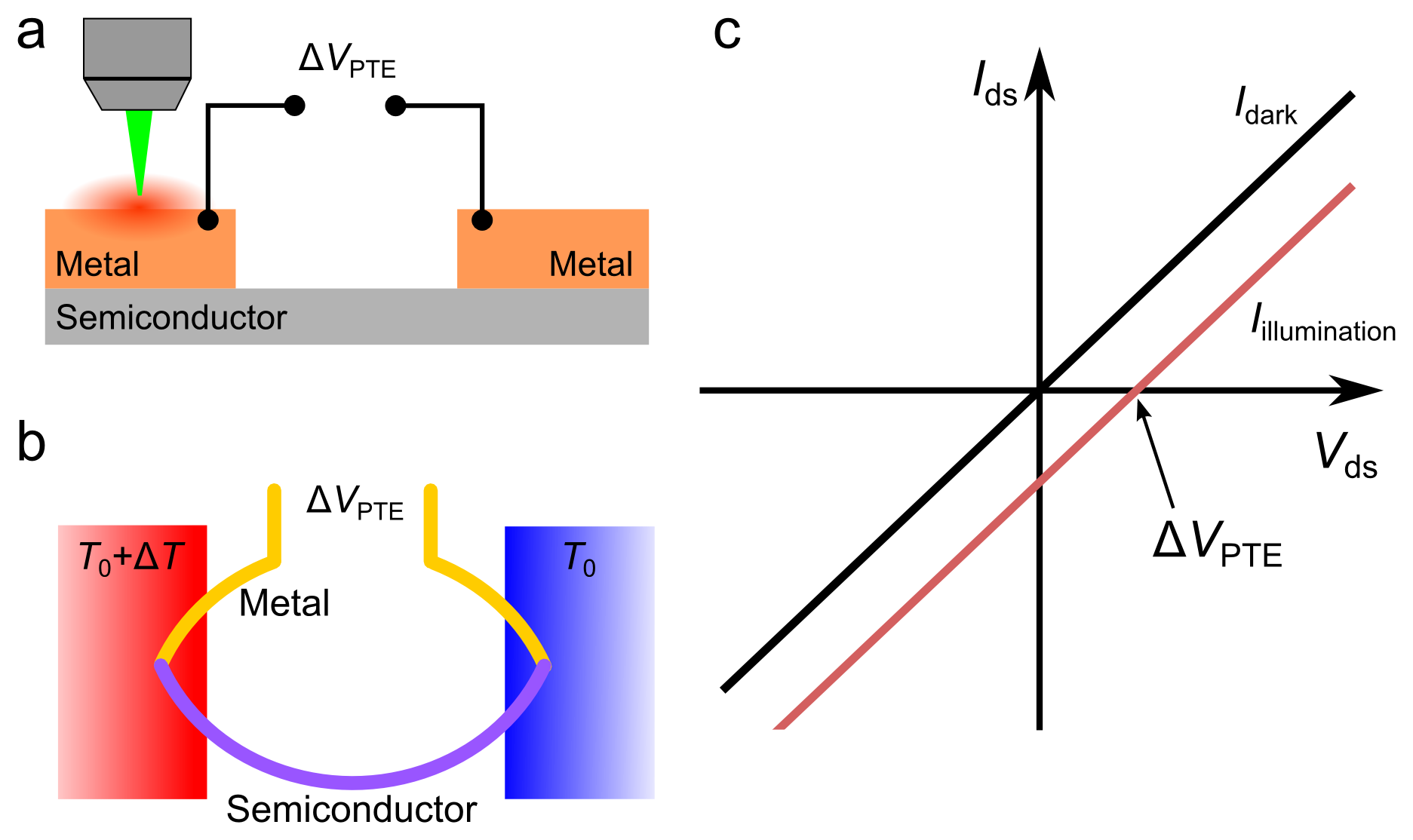}
\caption{Photo-thermoelectric effect. (a) Schematic of a field-effect transistor locally illuminated by a focused laser spot on one of the metal contacts to the semiconducting channel. The circuit is open and a thermoelectric voltage $\Delta V_\mathrm{PTE}$ develops across the contacts. (b) Thermal circuit corresponding to the device depicted in (a). The two metal/semiconductor junctions are kept at a steady-state temperature difference $\Delta T$, generating a photo-thermoelectric voltage $\Delta V_\mathrm{PTE}$. (c) \ids$-$\vds\ characteristics in the dark (black solid line) and under illumination (red solid line) of a device whose photoresponse is dominated by the photo-thermoelectric effect. Illumination results in a current at zero bias and a thermoelectric voltage $\Delta V_\mathrm{PTE}$ across the contacts when the circuit is kept open.}
\label{intro_seebeck2}
\end{figure}

In the photo-thermoelectric effect (PTE), a heat gradient from light-induced heating results in a temperature gradient across a semiconductor channel. As a result, the two ends of the semiconductor channel display a temperature difference $\Delta T$. This $\Delta T$ is converted into a voltage difference $\Delta V$ via the Seebeck (or thermoelectric) effect (see Figure \ref{intro_seebeck2}). The magnitude of $\Delta V$ is linearly proportional to the temperature gradient via the Seebeck coefficient ($S$):\cite{Ashcroft1976}
\begin{equation}\label{VSDT}
\Delta V = S \cdot \Delta T.
\end{equation}
The heat gradient can stem from either localized illumination, as with a focused laser spot with dimensions much smaller than those of the measured device,\cite{Park2009,Buscema2013} or from a strong difference in the absorption in distinct parts of the device under global illumination.\cite{Groenendijk2014}

The origin of the Seebeck effect can be found in three main microscopic processes, in dynamic equilibrium with each other.\cite{Donald1962, Ashcroft1976} Relating the Seebeck coefficient to microscopic quantities is, however, difficult. Therefore, the Seebeck coefficient is usually expressed in terms of the conductivity of the material through the Mott relation:\cite{Mott1936, Nolas2001, Bhush2007, Heremans2008}
\begin{equation}\label{Mott2}
S = \left.\frac{\pi^2 k_{\mathrm{B}}^2 T}{3e}\frac{d\ln(\sigma(E))}{dE}\right|_{E=E_F},
\end{equation}
where \kb\ is the Boltzmann constant and $\sigma(E)$ the conductivity as a function of energy ($E$), and where the derivative is evaluated at the Fermi energy, $E_F$. 
The sign of the Seebeck coefficient is determined by the majority charge carrier polarity in the semiconductor. \\

Through the photo-thermoelectric effect (PTE), a temperature gradient generates a voltage difference that can drive a current through a device at zero \vds. As sketched in Figure \ref{intro_seebeck2}, a device which is illuminated by a focused laser spot can be modeled as two junctions between the contact metal and the semiconductor channel. In this example, a focused illumination on one of the electrical contacts keeps a steady-state temperature difference ($\Delta T$) between the two junctions, leading to a voltage difference across them ($\Delta V_\mathrm{PTE}$):
\begin{equation}\label{intro_deltav_pte}
    \Delta V_\mathrm{PTE} = (S_\mathrm{semiconductor}-S_\mathrm{metal}) \cdot \Delta T \approx S_\mathrm{semiconductor} \cdot \Delta T.
\end{equation}
In Equation \ref{intro_deltav_pte}, the term $S_\mathrm{metal} \cdot \Delta T$ can usually be neglected because the Seebeck coefficients of pure metals are in the order of \SI{1}{\seebeck}, much smaller than typical values for semiconductors, reported in Table \ref{Seeb}. $\Delta T$ can be estimated via finite element simulations \cite{Slachter2010, Buscema2013,Li2014c} or measured with on-chip thermometers.\cite{Li2014c,Wu2014} Once $\Delta T$ is known, it is possible to estimate the Seebeck coefficient of the semiconductor material. Seebeck coefficient values for common materials are listed in Table \ref{Seeb}. The magnitude of $\Delta V_\mathrm{PTE}$ typically ranges from tens of \si{\micro\V} to tens of \si{\mV}. In order to drive current through the device, the electrodes metals need to form ohmic contacts to the semiconductor. In case the illumination is focussed on a uniform semiconducting channel, no current will flow in the device since no external bias is applied and negligible thermal gradients can be achieved.

\begin{table}

\centering

    \begin{tabular}{@{}lrc@{}}\toprule
     Material & Seebeck coefficient (\si{\seebeck})  &  Ref \\
     \midrule  
    Graphene & $\pm 4$ to $\pm 100$ & \cite{Xu2009,Gabor2011,Seol2010,Grosse2011,Wang2010a} \\
    CNT & $\sim 300$ & \cite{Small2003} \\
    Organic Semicond. & $1 \times 10^3$ & \cite{Pernstich2008} \\
    \ce{Bi2Te3} & $\pm 150$ & \cite{Fleurial1988,Bassi2009} \\
    \ce{MnO2} & $\sim 4 \times 10^4$ & \cite{Song2012} \\
    \ce{Fe2O3} & $\sim 1 \times 10^4$ &  \cite{Gardner1963} \\
    \ce{MoS2} & $ 1 \times 10^5$ &  \cite{Buscema2013} \\   
    \ce{MoS2} & $ 3 \times 10^4$ &  \cite{Wu2014} \\   
    Black phosphorus & $ 60$ &  \cite{Low2014} \\  
    Black phosphorus & $ 335$ &  \cite{Flores2015} \\   
    \bottomrule
    \end{tabular}
\caption{Seebeck coefficients of some thermoelectric materials. }
\label{Seeb}
\end{table}

Figure \ref{intro_seebeck2}c shows a typical \ids$-$\vds\ curve of a device where the photoresponse is dominated by the photo-thermoelectric effect. In the dark, the  \ids\ is linear with the \vds\, indicating that the Schottky barriers are small (ohmic contact). Under illumination, the photo-thermo\-e\-lec\-tric effect generates a current at zero bias, without changing the resistance of the device. $\Delta V_\mathrm{PTE}$ can be read off from the intersection with the zero-current axis, as illustrated in Figure \ref{intro_seebeck2}.

\subsubsection{Photo-bolometric effect.~~} 
The photo-bolometric effect (PBE) is based on the change of resistivity of a material due to uniform heating induced by photon absorption. The magnitude of this effect is proportional to the conductance variation of a material with temperature ($dG/dT$) and the homogeneous temperature increase ($\Delta T$) induced by laser heating. The bolometric effect underpins many detectors for mid and far-IR.\cite{Richards1994} Detectors working on this principle usually operate in a four-wire configuration to increase the accuracy in the determination of the conductance change.
Recent reports have also demonstrated the importance of the photo-bolometric effect for graphene-based detectors, where electron cooldown is limited by the available phonons,\cite{Freitag2013,Vicarelli2012} and for photodetectors based on \SI{\sim 100}{\nm} thick black phosphorus.\cite{Engel2014a, Low2014a}

\subsubsection{Comparison between photo-thermoelectric and bolometric effects.~~}

The main difference between the PTE and the PBE is that the PBE cannot drive currents in a device. Since the PBE is related to a change in the conductivity of the material with temperature, it will only modify the magnitude of the current flowing under external bias and illumination.

The sign of the photoresponse is also very different between the two mechanisms. In the PTE, the sign of the photocurrent is related to the difference in Seebeck coefficients between the components of the junction and, ultimately, to the type of charge carriers in the semiconductor material. For the PBE, the sign of the photocurrent is related to the change in the material conductivity with temperature.\cite{Richards1994}

From a device architecture perspective, the photo-thermoelectric effect requires a temperature difference between two junctions of materials with different Seebeck coefficients. In contrast, the photo-bolometric effect will be present in a homogeneous material, with a homogeneous temperature profile, and under external bias.

\section{Photodetectors figures-of-merit}\label{PD_fom}
To facilitate comparison between photodetectors working with different principles and realized with different materials and geometries, it is customary to use a set of figures-of-merit. In the following discussion, we will briefly introduce the general meaning of each figure of merit and give a reference to commercial silicon and \ce{InGaAs} photodetectors and also recent graphene-based detectors, as a benchmark. 

\paragraph*{Responsivity, $\mathfrak{R}$, measured in \si{\A\per\W}.~~} The responsivity is the ratio between the photocurrent and the total incident optical power on the photodetector. It is an indication of the achievable electrical signal under a certain illumination power: a large responsivity indicates a large electrical output signal for a defined optical excitation power. With the help of Equation \ref{photocurrent}, we can define the responsivity as: 
\begin{equation}\label{responsivity}
\begin{split}
\mathfrak{R} & = \frac{I_\mathrm{photo}}{P} = \Gamma \eta  e G \cdot \frac{\tau_\mathrm{photocarriers} \mu V}{L^2} \cdot \frac{\lambda}{\Gamma_\mathrm{total} h c} =\\
& = \frac{\Gamma e \lambda}{\Gamma_\mathrm{total} h c} \cdot \eta \cdot G \cdot \tau_\mathrm{photocarriers} \cdot \frac{\mu V}{L^2},
\end{split}
\end{equation}
where $\lambda$ is the photon wavelength, $h$ Planck's constant, $c$ the speed of light in vacuum and $\Gamma_\mathrm{total}$ is the total photon flux. Equation \ref{responsivity} shows that the responsivity still depends on the external bias ($V$) or geometrical factors, at least in the simplified model presented here. For commercial silicon photodiodes, the responsivity is in the order of \SI{500}{\respo} at a wavelength of \SI{880}{\nm} and becomes negligible for wavelengths shorter than \SI{405}{\nm} and larger than \SI{1100}{\nm}. Commercial \ce{InGaAs} detectors reach \SI{\sim 1.2}{\A\per\W} at $\lambda$ \SI{=1550}{\nm}. For graphene photodetectors, the responsivity is about \SI{10}{\respo} accross the visible and telecommunication wavelengths.\cite{Xia2009} Recently, more complex graphene detectors based on photogating either with quantum dots\cite{Konstantatos2012} or with another graphene sheet separated by a tunnel barrier\cite{Liu2014b} have achieved much larger responsivities: \SI{1e7}{\A\per\W} and \SI{1e3}{\A\per\W}, respectively.

\paragraph*{External quantum efficiency, $EQE$, dimensionless.~~} The external quantum efficiency is the ratio of the number of the charge carriers in the photocurrent $n_\mathrm{e}$ and the total number of impinging excitation photons $n^\mathrm{total}_\mathrm{photon}$. It is closely related to the responsivity $\mathfrak{R}$:
\begin{equation}\label{eqe}
EQE = \frac{n_\mathrm{e}}{n^\mathrm{total}_\mathrm{photon}} =  \mathfrak{R} \cdot \frac{hc}{e\lambda},
\end{equation}
The $EQE$ is a measure of the optical gain $G$ in the photodetector: $EQE > 1$ effectively means that more than one charge carrier per impinging photon is measured. It is usually a lower bound for the internal quantum efficiency since it assumes that all the incident photons are absorbed. For a silicon photodiode the EQE is in the order of \num{6e5} and for a graphene detector is about \num{1e-2}.\cite{Xia2009,Mueller2009} 

\paragraph*{Internal quantum efficiency, $IQE = \eta$, dimensionless.~~} The internal quantum efficiency is the number of measured charge carriers $n_\mathrm{e}$ divided by the number of \emph{absorbed} photons $n^\mathrm{abs}_\mathrm{photon}$. 

By accounting for the photon losses due to transmission and, for a small part, reflection, it is possible to determine the amount of absorbed photons and calculate the internal quantum efficiency. Compared to their bulk counterparts, ultrathin layered materials have an increased photon loss due to transmission. Therefore, at the same EQE, ultrathin layered materials have a larger IQE. To give a correct estimate of the absorbed photons, optical interference effects should be considered. These effects can enhance photon absorption up to 10\% $\sim$ 15\%.\cite{Blake2007,Casiraghi2007,Furchi2014a} 

\paragraph*{Time response, measured in \si{s}.~~} The time response of a photodetector is usually measured between 10\% and 90\% of the generated signal under modulated excitation intensity, either on the raising or falling edge. A photodetector with a small response time is usually desired to allow for certain applications, like video-rate imaging. Commercial silicon and \ce{InGaAs} photodiodes show rise times of about \SI{50}{\pico\s} while graphene detectors can already reach hundreds of picoseconds.\cite{Xia2009,Mueller2009} 

\paragraph*{Bandwidth, $B$, measured in \si{\Hz}.~~} The bandwidth of a photodetector is defined as the modulation frequency ($f_\mathrm{modulation}$) of the incoming light excitation at which the intensity of the detector signal is \SI{3}{\decibel} lower than under continuous illumination. A photodetector with a large bandwidth is desirable for high-rate optical information transfer. As for the response time, the bandwidth of commercial silicon and graphene detectors is similar and reaches a few tens \si{\GHz}.\cite{Xia2009,Mueller2009} Using optical correlation techniques, the intrinsic bandwidth of graphene has been estimated to be about \SI{260}{\GHz}.\cite{Urich2011} 

\paragraph*{Noise equivalent power, $NEP$, measured in $\mathrm{W Hz^{-\frac{1}{2}}}$.~~} The noise equivalent power is the minimum illumination power that delivers a unity signal-to-noise ratio at \SI{1}{\Hz} bandwidth. It is similar to the \emph{sensitivity} figure-of-merit for other type of detectors. It is a measure of the minimum detectable illumination power; thus a low $NEP$ is a desirable property of a photodetector. It can be estimated by:
\begin{equation}
NEP = \frac{PSD}{\mathfrak{R}},
\end{equation}
where the $PSD$ is the current noise power spectral density in dark (in $\mathrm{A Hz^{-\frac{1}{2}}}$) at \SI{1}{\Hz} bandwidth. Under the assumption of a shot-noise limited detector, the $PSD$ is proportional to the square root of the dark current. Thus a detector with a small dark current and large responsivity will have a small $NEP$. Commercial silicon photodiodes have a NEP in the order of \num{1e-14} $\mathrm{W Hz^{-\frac{1}{2}}}$, \ce{InGaAs} detectors reach one order of magnitude lower, and graphene detectors show values about \num{1e-12} $\mathrm{W Hz^{-\frac{1}{2}}}$, being limited by the high dark current. 

\paragraph*{Detectivity, $D^*$, measured in $\frac{\mathrm{cm}\sqrt{\mathrm{Hz}}}{\mathrm{W}}$}.~~ The detectivity ($D*$) is a figure-of-merit derived from the $NEP$, area and bandwidth that enables comparison between photodetectors of different geometries: a higher detectivity indicates a better photodetector performance. It is defined as:
\begin{equation}\label{detectivity}
D^* = \frac{\sqrt{AB}}{NEP}
\end{equation}
where $A$ is the area of the photodetector and $B$ is its bandwidth. The detectivity of silicon photodiodes is in the order of \num{1e12} $\frac{\mathrm{cm}\sqrt{\mathrm{Hz}}}{\mathrm{W}}$ while for hybrid graphene-\ce{PbS} systems can reach about \num{1e13} $\frac{\mathrm{cm}\sqrt{\mathrm{Hz}}}{\mathrm{W}}$, similar to commercial detectors based on III-V materials.\cite{Konstantatos2012} 

\paragraph*{Wavelength range, measured in \si{\nm}.~~} The wavelength range indicates which part of the electromagnetic spectrum can be detected, at a given $NEP$. The limiting factor is usually the bandgap of the active material, which determines the absorption edge. For the detectors relying on photons to generate an electron-hole pair, the absorption edge defines the longest wavelength that can be detected. Detectors working with thermal processes can overcome this limit.

\section{Transition metal (\ce{Mo},\ce{W}) dichalcogenides photodetectors}

\begin{figure*}[t]
\captionsetup{width=.9\textwidth}
\centering
\includegraphics[]{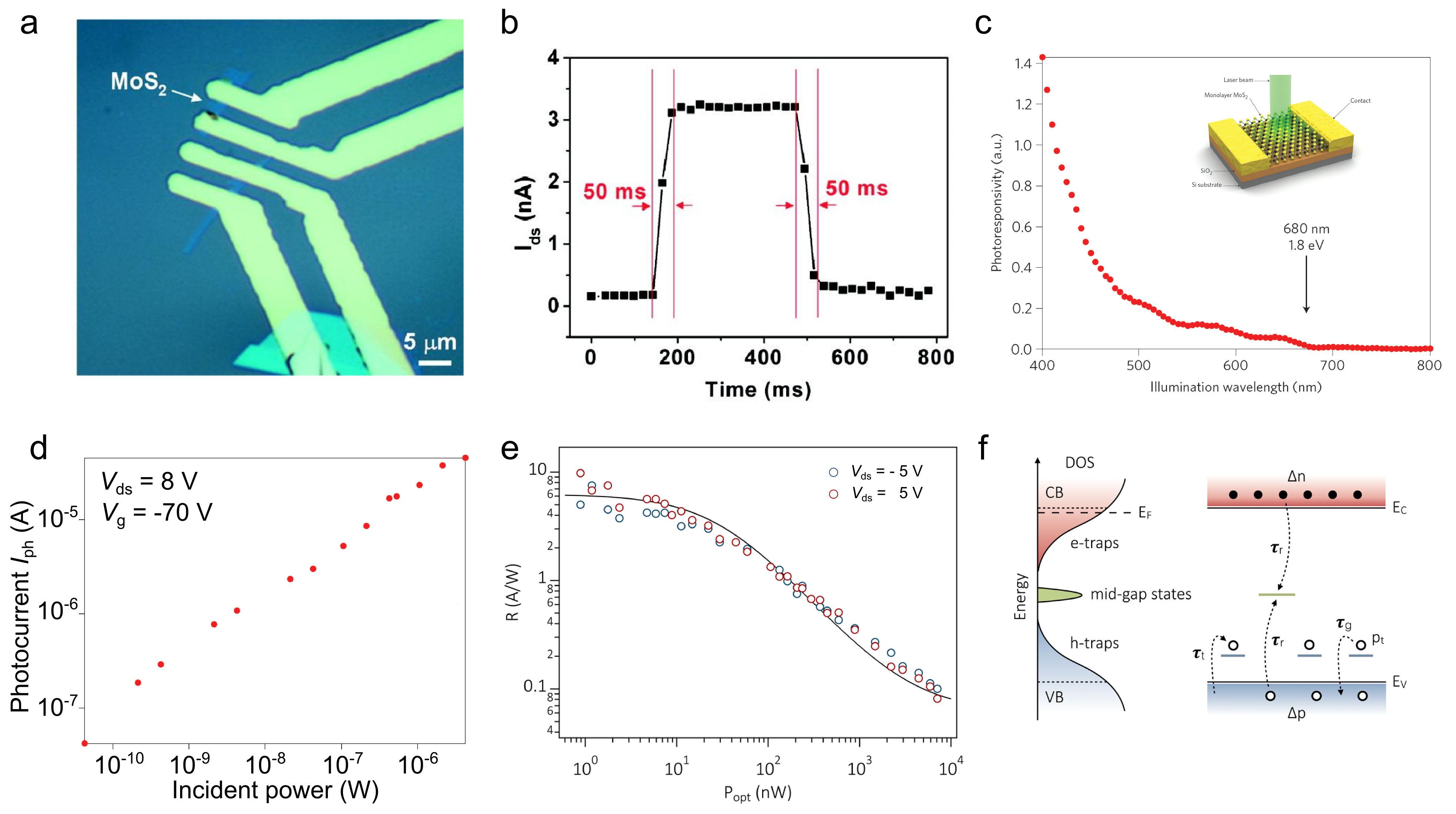}
\caption{(a) Optical image of a typical device based on single layer \ce{MoS2}. (b) Time response of the device in panel (a). (c) Photoresponsivity as a function of illumination wavelength, evidencing the cut-off wavelength at \SI{680}{\nm}. Inset: device schematics. (d) Photocurrent as a function of incident optical power. (e) Responsivity as a function of excitation power for \vds\ \SI{=-5}{\V} (blue dots) and \vds\ \SI{=5}{\V} (red dots). Solid black line: fit with parameters from panel f. (f) Energy and relevant rates for the trap-assisted recombination model. Hole traps are the only relevant traps due to the large effective mass of the holes (open dots) compared to the electrons (closed dots) . Panels a,b are adapted with permission from ref. \cite{Yin2012} copyright 2012 American Chemical Society. Panels c,d are adapted by permission from Macmillan Publishers LTD: Nature Nanotechnology ref. \cite{Lopez-Sanchez2013}, copyright 2013,. Panels e,f adapted with permission from ref. \cite{Furchi2014} copyright 2014 American Chemical Society.}
\label{materials_mos2_4}
\end{figure*}
Transition metal dichalcogenides (TMDCs) are layered compounds with general formula \ce{MX2} where \ce{M} is a metal from group IV, V or VI of the transition metals and \ce{X} is a chalcogen atom (group VI such as \ce{S}, \ce{Se} or \ce{Te}).\cite{Wang2012} In every layer, the metal atoms are covalently bonded to the chalcogen atoms on either side (for a schematic, see Figure 12). This covalent bond provides the structural integrity to a single layer. To form a bulk crystal, several layers are held together by weak van der Waals interaction. This weak out-of-plane interaction facilitates micromechanical exfoliation, allowing isolation of single layers. The mechanical exfoliation of these materials has opened the door to the fabrication of FETs based on single- and few-layer TMDCs and the study of their (opto)electronic properties. For more details on TMDCs, including their structure and their electronic properties, we refer the reader to refs.\cite{Wang2012,Fiori2014}

\subsection{Molybdenum disulphide} 
\ce{MoS2} is the most studied semiconducting TMDC. Its large and direct bandgap (\SI{1.8}{\eV})\cite{Ayari2007,Mak2010,Splendiani2010,Eda2011}, its mobility\cite{Radisavljevic2011,Radisavljevic2013} above \SI{100}{\mobility} and its remarkable mechanical properties\cite{Bertolazzi2011,Castellanos-Gomez2012} make single and few-layer \ce{MoS2} an interesting material for optoelectronic and flexible devices.\cite{Pu2012,He2012,Yin2012,Lee2012,Choi2012,Lopez-Sanchez2013,Buscema2013,Fontana2013,Tsai2013,Wu2013, Zhang2013a, Furchi2014}

\paragraph*{Externally biased photodetectors.~~}
Photodetectors based on single- and few-layer \ce{MoS2} are usually in the form of photo-FETs (see Figure \ref{materials_mos2_4}a). Yin \emph{et al.}\cite{Yin2012} first reported a photo-FET based on single layer \ce{MoS2} (Figure \ref{materials_mos2_4}a) reaching a responsivity of about \SI{7.5}{\respo} in electron accumulation, measurable photoresponse up to about \SI{750}{\nm} and a response time in the order of \SI{50}{\ms} (see Figure \ref{materials_mos2_4}b).\cite{Yin2012} 

Recently, a similar study by Lopez-Sanchez \emph{et al.}\cite{Lopez-Sanchez2013} reported a low-bound responsivity for exfoliated 1L \ce{MoS2} photo-FET of \SI{880}{\A\per\W} in depletion and a cut-off wavelength of \SI{680}{\nm} (see Figure \ref{materials_mos2_4}c).\cite{Lopez-Sanchez2013} The reported responsivity is a lower bound since it is estimated in the OFF state of the photo-FET, where the photocurrent magnitude is the lowest.\cite{Lopez-Sanchez2013} With increasing power, the photocurrent raises sublinearly (Figure \ref{materials_mos2_4}d). The time response is in the order of \SI{4}{\s} and it could be reduced to \SI{600}{\ms} with the help of a gate pulse to reset the conductivity of the FET.\cite{Lopez-Sanchez2013} Both the sublinear behavior of the photocurrent with power and the need of a gate pulse to reset the conductivity are symptomatic of a photocurrent generation mechanism where trap states play a dominant role.

The NEP reached by the photodetector presented by Lopez-Sanchez \emph{et al.} is \num{1.8e-15} $\mathrm{W Hz^{-\frac{1}{2}}}$.\cite{Lopez-Sanchez2013} This remarkably low NEP stems from the low magnitude of the dark current in electron depletion, achievable due to efficient field-effect tunability in single-layer \ce{MoS2} and the \SI{1.8}{\eV} bandgap of 1L \ce{MoS2} that suppresses thermally activated carriers.\cite{Radisavljevic2011}

In the case of 1L \ce{MoS2} grown by chemical vapor deposition (CVD), Zhang \emph{et al.} reported responsivities in the order of \SI{2200}{\A\per\W} in vacuum and \SI{780}{\A\per\W} in air, highlighting the importance of the environment for both the electronic and the optoelectronic properties.\cite{Zhang2013a} Given the large surface-to-volume ratio, adsorbates play a major role in the properties of layered materials. \cite{Schedin2007, Fang2012, Late2012a,Tongay2013} Charged adsorbates also reduce the photoresponse as they suppress the lifetime of trapped carriers by acting as recombination centers.\cite{Zhang2013a} In a similar study on CVD grown \ce{MoS2}, Perea-L\'opez \emph{et al.}\cite{Perea-Lopez2014} found a responsivity in the order of \SI{1.1}{\respo}, five orders of magnitude lower than in the case of Zhang \emph{et al.}.\cite{Zhang2013a}

There are a few likely reasons for the large difference in responsivity. At first, there is a difference in device resistance. The device measured by Zhang \emph{et al}.\cite{Zhang2013a} is subjected to a laser annealing procedure in vacuum, which reduces the device resistance. Moreover, the measurements in ref \cite{Zhang2013a} are performed with a large and positive gate voltage, which brings the device in the ON state. The resulting device resistance is about \SI{20}{\kilo\ohm}. For comparison, the device measured by Perea-L\'opez \emph{et al}.\cite{Perea-Lopez2014} has a resistance of about \SI{2}{\giga\ohm} (no annealing and zero gate voltage).
Another important reason is the interdigitated electrode geometry of the device measured by Zhang \emph{et al}.\cite{Zhang2013a}. The interdigitated device geometry grants a much (about 4 orders of magnitude) larger active area which contributes to a larger photocurrent and, therefore, a larger responsivity. Also the difference in spot size plays a role. Zhang \emph{et al}.\cite{Zhang2013a} employ a laser spot of about \SI{1}{\mm} diameter, which allows them to expose the whole active area of the detector. In the study of Perea-lopez \emph{et al}.\cite{Perea-Lopez2014}, a \SI{2}{\um} laser spot is used, limiting the illuminated area to a small portion of the device.
Finally, as a consequence of the larger active area and lower resistance, Zhang \emph{et al}.\cite{Zhang2013a} are able to illuminate the device with lower optical power before hitting the noise floor of the current amplifier. Usually, at low optical powers the responsivity is the highest (see Equation 8). Therefore, the applied gate voltage, the interdigitated electrode geometry and the lower optical excitation power all contribute to the much larger responsivity reported by Zhang \emph{et al}.\cite{Zhang2013a}
 
Both exfoliated and CVD-grown \ce{MoS2} display large responsivity (large optical gain), sub-linear \iphoto\ with incident optical power and \si{ms}-to-\si{\s} response times; all these properties point to a photogeneration mechanism in which trap states play a major role. In particular, the photocurrent generation mechanism is a combination of both photoconductance (PC) and photogating (PG), as recently reported in ref \cite{Furchi2014}. Furchi \emph{et al.}\cite{Furchi2014} study the photoresponse of single and bilayer \ce{MoS2} photo-FETs by varying the incident optical power and the modulation frequency of the optical excitation. The large difference in time scales between PC (fast) and PG (slow) facilitates disentangling the two effects. 

The slow PG is attributed to long-lived trap states at the interface between the \ce{MoS2} channel and the \ce{SiO2} surface, likely hydrated by water adsorbates. The surface density of these trap sites is estimated to be in the order of \SI{1e15}{\per\square\cm} from DC photocurrent measurements, in agreement with previous literature on carbon-nanotube transistors.\cite{Kim2003} These \ce{SiO2}-bound trap states are likely to play a role in the photoresponse of other unencapsulated devices built on \ce{SiO2} and measured in ambient condition or even in vacuum without in-situ annealing.
The fast PC response is attributed to both mid-gap states and hole-trap states that are caused by defects within the \ce{MoS2} layer. These types of states have a faster decay rate compared to the \ce{SiO2}-bound trap states and thus result in a lower optical gain. Furchi \emph{et al}.\cite{Furchi2014} measure the PC response by a lock-in technique, modulating the light excitation at a frequency faster than the PG effect. To model the acquired data (Figure \ref{materials_mos2_4}e), a model based on mid-gap states and hole-trap states is proposed (Figure \ref{materials_mos2_4}f). From this model they extract the density of hole-traps to be about \SI{5e10}{\per\square\cm}, in good agreement with values obtained by independent transport measurements.\cite{Ghatak2013}

\paragraph*{Multilayer \ce{MoS2} photodetectors.~~}
Multilayer \ce{MoS2} photodetectors benefit from the bandgap reduction, allowing for extended detection range,\cite{Choi2012, Lee2012} larger absorption due to increased thickness,\cite{Wu2013} and achieve responsivity much lower than the best single-layer photodetectors. The reduction in responsivity is possibly due to their indirect bandgap.\cite{Tsai2014, Furchi2014} The photodetection mechanism in multilayer \ce{MoS2} is still dominated by trap states.

In conclusion, biased photodectors based on single- and few-layer \ce{MoS2} routinely achieve high responsivity, indicating high photogain, originating from the long lifetime of photogenerated carriers, boosted by trap states. Table \ref{materials_mos2_table} summarizes the main figures-of-merit of this type of photodetectors.

\paragraph*{Photovoltaic and photo-thermoelectric effects in \ce{MoS2}.~~}

\begin{figure}[h!]
\captionsetup{width=.9\textwidth}
\centering
\includegraphics[width = .8\textwidth]{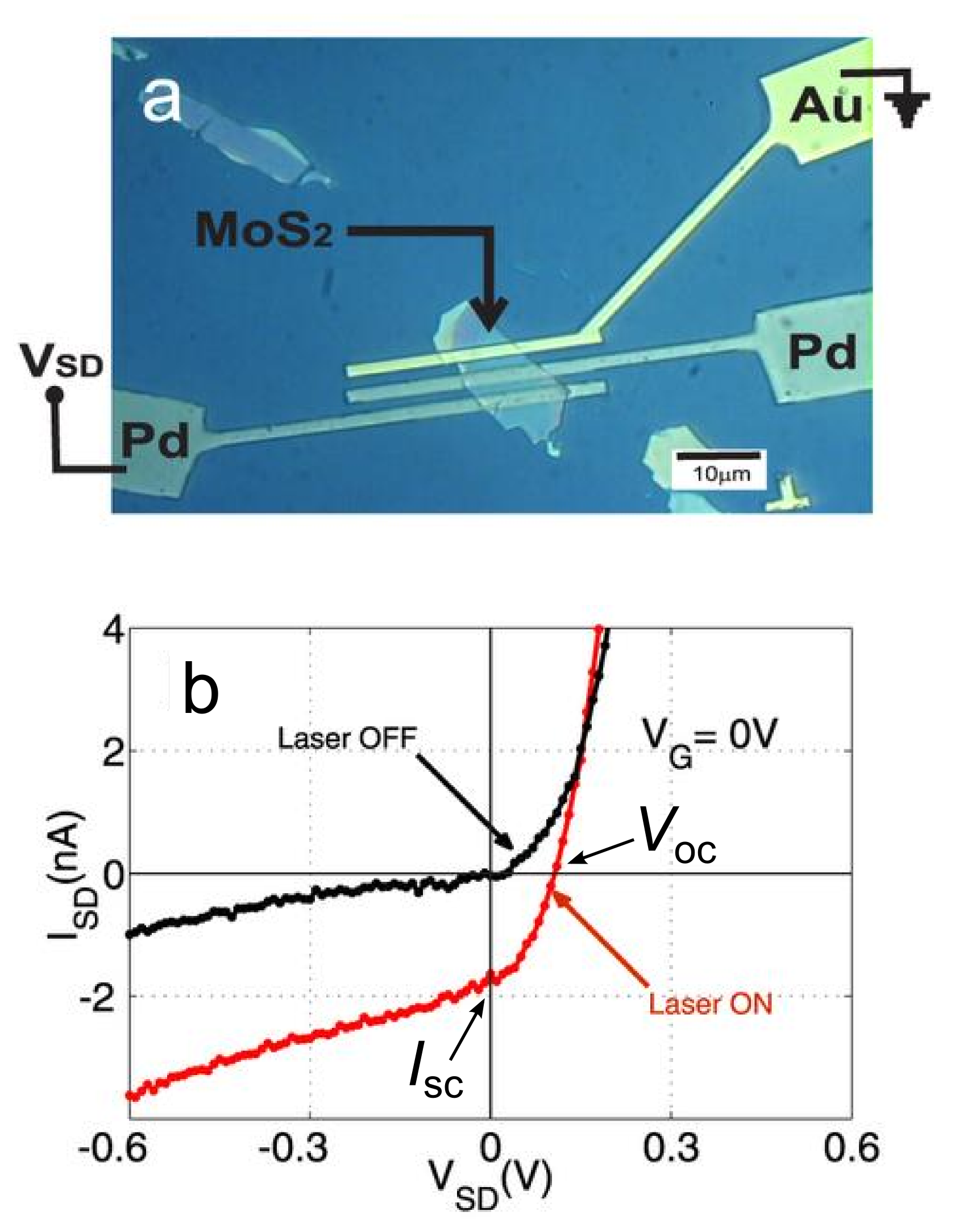}
\caption{(a) Optical image and circuit schematic of a typical ultrathin \ce{MoS2} device contacted with different metals (\ce{Au} and \ce{Pd}). (b) \ids$-$\vds\ characteristics of the device in panel a, in dark (solid black line) and under illumination (solid red line). Both panels are adapted by permission from Macmillan Publishers LTD: Scientific Reports ref. \cite{Fontana2013}, copyright 2013 }
\label{materials_mos2_6}
\end{figure}

In a recent study, Fontana \emph{et al.}\cite{Fontana2013} reported photovoltaic effect in a photo-FET based on a thin (\SI{\sim 50}{\nm}) \ce{MoS2} flake where the source/drain contacts were made from gold and palladium, respectively.\cite{Fontana2013} The workfunction difference between \ce{Au} and \ce{Pd} promotes a small difference in local doping of the \ce{MoS2} flake, giving rise to an asymmetry in the SBs at the two contacts, as in the case of graphene phototransistors.\cite{Fontana2013, Mueller2009} The SBs asymmetry effectively acts as an internal electric field and generates a photocurrent under global illumination ($r_\mathrm{spot} \gg L_\mathrm{device}$). An independent study on a similar device based on single-layer \ce{MoS2} with \ce{Au} and \ce{Pd} contacts has demonstrated that the SB height for \ce{Pd} contacts is larger than for \ce{Au}.\cite{Kaushik2014}

Figure \ref{materials_mos2_6}a shows a micrograph of the device by Fontana \emph{et al.},\cite{Fontana2013} highlighting the asymmetric metallization of the contacts. The measured \ids$-$\vds\ characteristics are plotted in Figure \ref{materials_mos2_6}b and show diode-like behavior in the dark and a clear PV photocurrent generation under illumination. For different devices, the short-circuit current $\isc$\ ranges from \SIrange[range-units = single]{0.2}{2}{\nA} and the open-circuit voltage $\voc$\ from about \SIrange[range-units = single]{30}{100}{\mV}. The large current in reverse bias indicates non-ideal diode behavior.

\begin{figure}[h!]
\captionsetup{width=.9\textwidth}
\centering
\includegraphics[width = 0.9\textwidth]{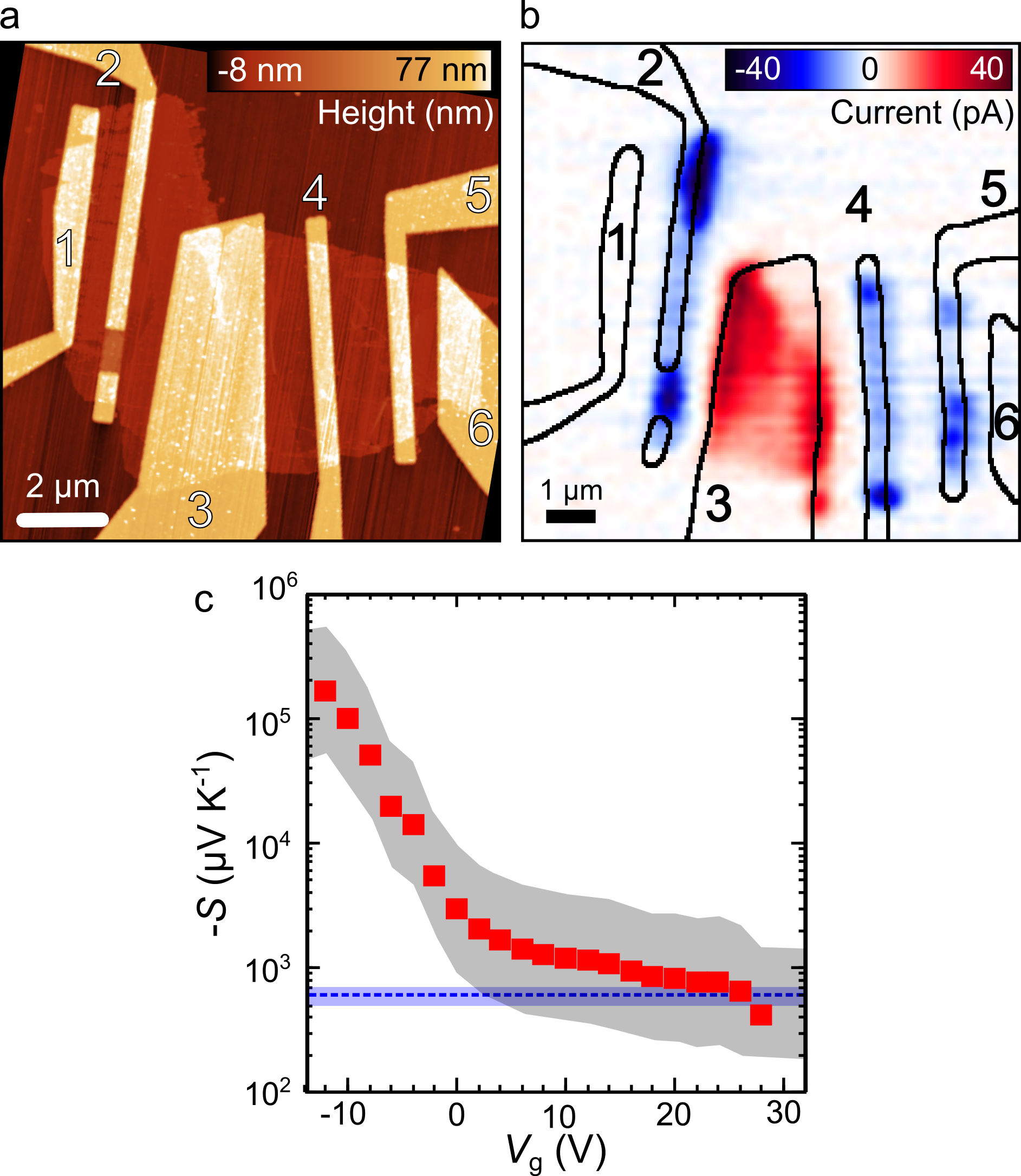}
\caption{(a) AFM image of one of the studied devices. Scale bar is \SI{2}{\um}. (b) SPCM map. The photocurrent (linear colorscale) is collected simultaneously with the intensity of the reflected light, which is used to detect the contour of the electrodes (solid black lines). Excitation is provided by a CW laser, $\lambda$ \SI{=532}{\nm}, $P$ \SI{=1}{\uW}, spot radius $\sim$ \SI{400}{\nm}. (c) Estimated Seebeck coefficient vs. gate voltage.  The gray shaded area is the uncertainty in $S$ due to the uncertainty in the estimation of the temperature difference. The dashed light blue line corresponds to the Seebeck coefficient value of bulk \ce{MoS2} with experimental uncertainty (shaded light blue area). All panels are adapted with permission from ref. \cite{Buscema2013} copyright 2013 American Chemical Society. }
\label{my_mos2}
\end{figure}

Buscema \emph{et al.}\cite{Buscema2013} have performed a scanning photocurrent microscopy (SPCM) study of a single layer \ce{MoS2} photo-FET and found a strong and tunable photo-thermoelectric effect.\cite{Buscema2013}  Figure \ref{my_mos2}a shows an AFM image of one of the studied devices. The contacts (\ce{Ti}/\ce{Au}) are deposited on top of an exfoliated \ce{MoS2} flake on a \ce{SiO2}/\ce{Si} substrate. The flake is single-layered between contacts 3 and 6. Electrode 3 is connected to a current-to-voltage amplifier while the other electrodes are grounded. 

With above-bandgap illumination, the SPCM measurements shown in Figure \ref{my_mos2}b reveal that the photocurrent is generated even 
when the laser spot is focused \emph{on top of} the electrode area, as particularly evident for electrode 3. Measurements performed with below bandgap illumination show the same qualitative behavior, with lower photocurrent values.\cite{Buscema2013} The current generation inside the electrode area and its persistence with below-bandgap illumination point to a photocurrent generation mechanism dominated by the photo-thermoelectric effect, as already demonstrated for graphene also for ultrafast terahertz detection.\cite{Xia2009a,Cai2014} From the sign of the photocurrent, the authors concluded that Seebeck coefficient of single-layer \ce{MoS2} is negative, as expected for an n-type semiconductor. 

The magnitude of the Seebeck coefficient $S$ can be calculated by measuring the generated photo-thermoelectric voltage $V_\mathrm{PTE}$ and dividing it by the temperature gradient, estimated via finite element simulations. The resulting $S$ is plotted against the gate voltage in Figure \ref{my_mos2}c. In electron accumulation, the $S$ of single-layer \ce{MoS2} is in agreement with its bulk value. As the channel is depleted, the Seebeck coefficient rapidly increases by two orders of magnitude, reaching \SI{\sim 1e5}{\seebeck}. This value is in reasonable agreement with recent electrical measurements where the temperature gradient is monitored with on-chip heaters and thermometers.\cite{Wu2014} This large value of $S$ and its gate tunability render single-layer \ce{MoS2} an interesting material for applications such as thermoelectric nanodevices.

We note that, under external bias and focussed illumination, also the PV effect at the Schottky barriers (SBs) could play a role. This has, for example, been reported in multilayer \ce{MoS2} devices which have shown photocurrent generation from SBs under external bias.\cite{Wu2013}

In conclusion, zero-bias photocurrent generation in single-layer \ce{MoS2} is dominated by the PTE effect due to the large values of the $S$ coefficient, while, in multilayer samples the PV effect can significantly contribute to the photocurrent.

\subsection{Molybdenum diselenide} 
Molybdenum diselenide (\ce{MoSe2}) has been grown via CVD techniques by several groups,\cite{Boscher2006,Chang2014,Lu2014,Shaw2014,Shim2014,Utama2014,Wang2014,Xia2014a} yielding large-area triangular flakes or continuous films on a variety of substrates. Photodetectors were patterned on the as-grown films or flakes were transferred to an oxidized silicon wafer to fabricate photo-FETs.

Typical CVD grown \ce{MoSe2} photodetectors show responsivity between \SI{0.26}{\respo}(ref.\cite{Chang2014}) and \SI{13}{\respo} (ref. \cite{Xia2014a}). Recently, a photodetector based on a \SI{\sim20}{\nm} thick \ce{MoSe2} exfoliated flake, deterministically transferred on \ce{Ti} electrodes, demonstrated a responsivity\cite{Abderrahmane2014} up to \SI{97.1}{\A\per\W}. The response time is in the order of few tens of milliseconds for all the studied devices. 

\begin{figure*}[t]
\captionsetup{width=.9\textwidth}
\centering
\includegraphics[width = 0.8\textwidth]{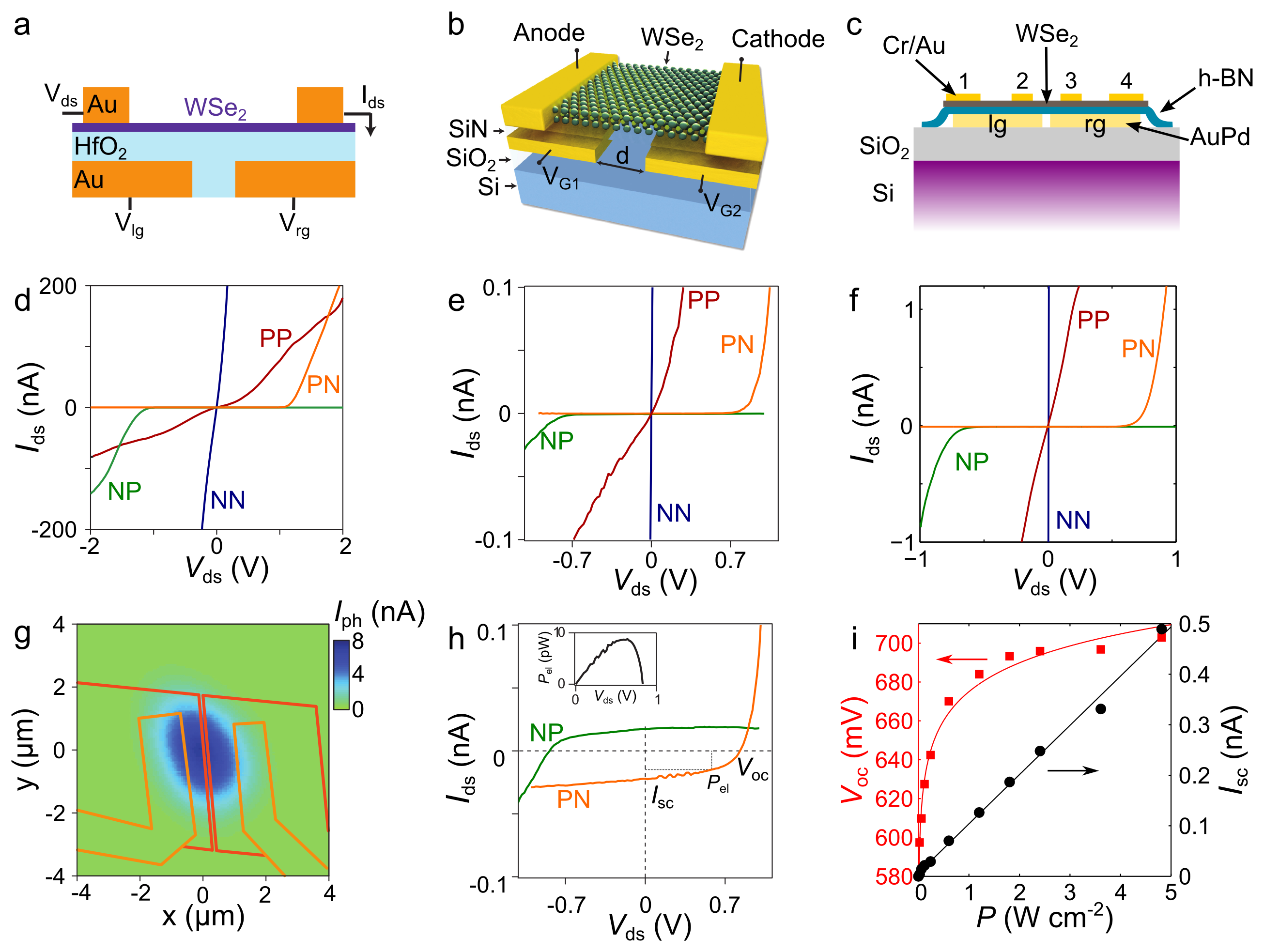}
\caption{(a) to (c) Device schematics: panel (a) from ref. \cite{Baugher2014}, panel (b) from ref. \cite{Pospischil2014} and panel (c) from ref. \cite{Groenendijk2014}. (d) to (f) \ids$-$\vds\ characteristics measured in different gate configurations; (d) from ref.\cite{Baugher2014}, (e) from \cite{Pospischil2014} and (f) from ref. \cite{Groenendijk2014}. (g) SPCM map of the photocurrent of the device in (a) in PN configuration. (h) \ids$-$\vds\ characteristics measured in PN and NP configuration under white light illumination for the device in (b). Inset: electrical power ($P_\mathrm{el}$) extracted from the device. (i) $\voc$ (left axis) and $\isc$ (right axis) as a function of excitation power for the device in (c). Panels (a), (d) and (g) are adapted by permission from Macmillan Publishers LTD: Nature Nanotechnology, reference \cite{Baugher2014}, copyright 2014. Panel (b),(e) and (h) are adapted by permission from Macmillan Publishers LTD: Nature Nanotechnology, reference \cite{Pospischil2014}, copyright 2014. Panels (c),(f) and (i) are adapted with permission from ref. \cite{Groenendijk2014}, copyright 2014, American Chemical Society. For the \ids$-$\vds\ in panels (d) to (f) the gate configurations are as follows: PP (both gates with negative bias), PN and NP (gate at opposite bias), NN (both gates at positive bias). The specific values of the gates voltages differ per device. }
\label{materials_wse_1}
\end{figure*}

\subsection{Tungsten disulfide}
Tungsten disulfide (\ce{WS2}) has also been employed for photodetection.\cite{Perea-Lopez2013, Huo2014} In their recent study, Huo \emph{et al.}\cite{Huo2014} show a significant dependence of the photoresponse on the gaseous environment in which the measurements were performed. \cite{Huo2014} At low excitation powers, the responsivity ranges from \SI{\sim 13}{\A\per\W} in vacuum to \SI{884}{\A\per\W} in a \ce{NH3} atmosphere. The increase of responsivity in the \ce{NH3} atmosphere was attributed to charge transfer from the absorbed molecules to the \ce{WS2} flake, which affects the \ce{WS2} doping level, as seen in luminescence experiments.\cite{Tongay2013} As a result of the charge transfer, the lifetime of (one of the) photogenerated carriers can be extended, leading to an enhancement in the responsivity.

The relevant figures-of-merit for \ce{MoSe2} and \ce{WS2} photodetectors are summarized in Table \ref{materials_mos2_table} to facilitate a direct comparison between photodetectors based on \ce{MoS2}, \ce{MoSe2}, \ce{WS2} and \ce{WSe2}.

\subsection{Tungsten diselenide} 

\paragraph*{Photodetectors.~~} Zhang \emph{et al.}\cite{Zhang2014} demonstrate the large impact of the metal contacts to a tungsten diselenide (\ce{WSe2}) flake on the photoresponse. The responsivity reaches \SI{180}{\A\per\W} with \ce{Pd} contacts and decreases by a factor $\sim 30$ with \ce{Ti} contacts. Conversely, the time response for the devices with \ce{Ti} contacts is less than \SI{23}{\ms} while for the \ce{Pd} contacts it is in the order of tens of seconds. The large variation in responsivity is attributed to the large difference in SBs induced by \ce{Pd} and \ce{Ti}, which is of general importance for photodetectors based on 2D materials.

\paragraph*{Electrostatically defined PN junctions.~~} In contrast to \ce{MoS2}, \ce{MoSe2} and \ce{WS2}, devices based on \ce{WSe2} readily demonstrate ambipolar transport via electrostatic gating.\cite{Podzorov2004,Das2013,Huang2013,Tosun2014,Chuang2014,Allain2014} Exploiting this property, locally-gated PN junctions have been realized with single-layer \ce{WSe2}.\cite{Baugher2014,Pospischil2014,Ross2014,Groenendijk2014} In the following, we will review the recent studies concerning such PN junctions. We refer the reader to Table \ref{materials_mos2_table} for a synthetic presentation of the photodetection performance of both the photodetector of ref.\cite{Zhang2014} and the PN junctions.

The schematics of the locally-gated PN junction devices in references \cite{Baugher2014,Pospischil2014,Groenendijk2014} are presented in Figures \ref{materials_wse_1}a,b,c. The fabrication of these PN junctions relies on the deterministic transfer of single-layer \ce{WSe2} on top of two local gates covered with a dielectric, which can be made from a conventional (\ce{HfO2} \cite{Baugher2014} or \ce{Si3N4} \cite{Pospischil2014}) or layered (\ce{hBN})\cite{Ross2014,Groenendijk2014} material. The single-layer \ce{WSe2} is then contacted with electrodes providing low Schottky barriers, usually a \ce{Cr}/\ce{Au} stack.\cite{Baugher2014, Groenendijk2014}

The local gates allow effective control of the charge carrier type and density in the \ce{WSe2} channel and, thus, can induce different carrier types in adjacent parts of the channel, giving rise to a PN junction. Looking at the \ids$-$\vds\ characteristics highlights the effect of the local gates on the electrical transport, as Figures \ref{materials_wse_1}d,e,f illustrate. As a function of the gate configuration, the \ids$-$\vds\ curves range from metallic (PP or NN) to rectifying with opposite direction (PN or NP).

Consistently accross three independent studies, the \ids$-$\vds\ curves in PN and NP configuration present a high resistance in parallel to the device ($R_\mathrm{par}$) and a low saturation current ($I_\mathrm{sat}$), both indicating nearly ideal diode behavior. This favorable behavior can be attributed to the large bandgap of single-layer \ce{WSe2} which reduces parallel pathways for current flow, such as thermally-activated carriers and field emission. The devices are also less resistive in the NN configuration than in the PP configuration. This may indicate a larger SB for hole injection resulting from the \ce{Cr}/\ce{Au} contacts. The difference in the absolute magnitude of \ids\ between these studies may be attributed to differences in device fabrication, gate-induced carrier density or measurement conditions.

We now turn our attention to the response to light excitation of such locally gated PN junctions. Figure \ref{materials_wse_1}g shows a false-color map of the photocurrent obtained by SPCM on the device in Figure \ref{materials_wse_1}a in NP gate configuration.\cite{Baugher2014} Even though the spatial resolution is limited by the long wavelength of the excitation laser, the photocurrent peaks at the depletion region (separation between the gates), suggesting that the photocurrent generation is indeed dominated by the PN junction electric field.

In Figure \ref{materials_wse_1}h we show the \ids$-$\vds\ in PN and NP configuration under illumination for the device in Figure \ref{materials_wse_1}b. In both curves, the reverse current increases, giving rise to a short-circuit current $\isc$ and an open-circuit voltage $\voc$. These signatures clearly indicate that the photocurrent generation is driven by the photovoltaic effect, as expected in a PN junction. The extracted electrical power is between tens of \si{\pico\W}\cite{Pospischil2014} and hundreds of \si{\pico\W} (see inset of Figure \ref{materials_wse_1}h for a representative plot).\cite{Groenendijk2014,Baugher2014} 

Figure \ref{materials_wse_1}i plots the $\voc$ (left axis) and $\isc$ (right axis) for increasing excitation power. The $\isc$ increases linearly while the $\voc$ increases logarithmically, confirming the ideal diode behavior, observed in the three independent studies.\cite{Baugher2014, Pospischil2014, Groenendijk2014} As a zero-biased photodetector, the external quantum efficiency (EQE) can be estimated from $\isc$ and it is in the order of $0.1 \sim 0.2$\% at $\lambda$\ \SI{=532}{\nm} (see Figure \ref{materials_wse_1}f, left axis).\cite{Baugher2014,Groenendijk2014} The NEP of detectors based on locally gated single-layer \ce{WSe2} is expected to be low in photovoltaic operation, due to the negligible dark current. The time response is in the order of \SI{10}{ms} and reduces in reverse bias.\cite{Groenendijk2014}

\section{\ce{Ga}, \ce{In} and \ce{Sn} chalcogenide photodetectors}\label{sec_IV}

Besides the previously discussed chalcogenides, also gallium $-$ \ce{Ga} $-$ and indium $-$ \ce{In} $-$ chalcogenide compounds are interesting for application as photodetectors.\cite{Hu2012,Late2012b,Liu2012,Hu2013,Jacobs-Gedrim2013,Mudd2013,Lei2014,Tamalampudi2014} Similarly to the TMDCs, these layered materials are held together with van der Waals forces. However, the structure of a single layer is different from single-layered TMDCs. In compounds with 1:1 stoichiometry, like \ce{GaS}, the atoms in each layer have the following repeating unit: \ce{S}$-$\ce{Ga}$-$\ce{Ga}$-$\ce{S} (Figure \ref{fig_dev_comparison}c). 
In a direction perpendicular to the layer plane, the atoms are arranged in a graphene-like honeycomb lattice.\cite{Hu2013} For \ce{GaS} and \ce{GaSe}, this crystal structure leads to a band structure with dominant indirect transitions (indirect: \SI{2.59}{\eV}, direct \SI{3.05}{\eV} for \ce{GaS}; indirect: \SI{2.11}{\eV}, direct \SI{2.13}{\eV} for \ce{GaSe}).\cite{Capozzi1989,Alekperov1991,Genut1992,Plucinski2003, Ho2006} The structure of \ce{GaTe} is slightly different, leading to a dominant direct transition (\SI{1.7}{\eV}). \cite{Sanchez-Royo2002} Compounds with different stoichiometry (like \ce{In2Se3}) have a more complex layer structure (Figure \ref{fig_dev_comparison}c) and a variety of structural phases influencing their electrical properties. The $\alpha$-phase has a direct bandgap of \SI{1.3}{\eV}.\cite{Groot2001,Sanchez-Royo2001,Sreekumar2008} The numerical values of the bandagap are summarized in Table \ref{bgap_tri_III} for convenience.

\begin{figure}[h!]
\captionsetup{width=.9\textwidth}
\centering
\includegraphics[width = \textwidth]{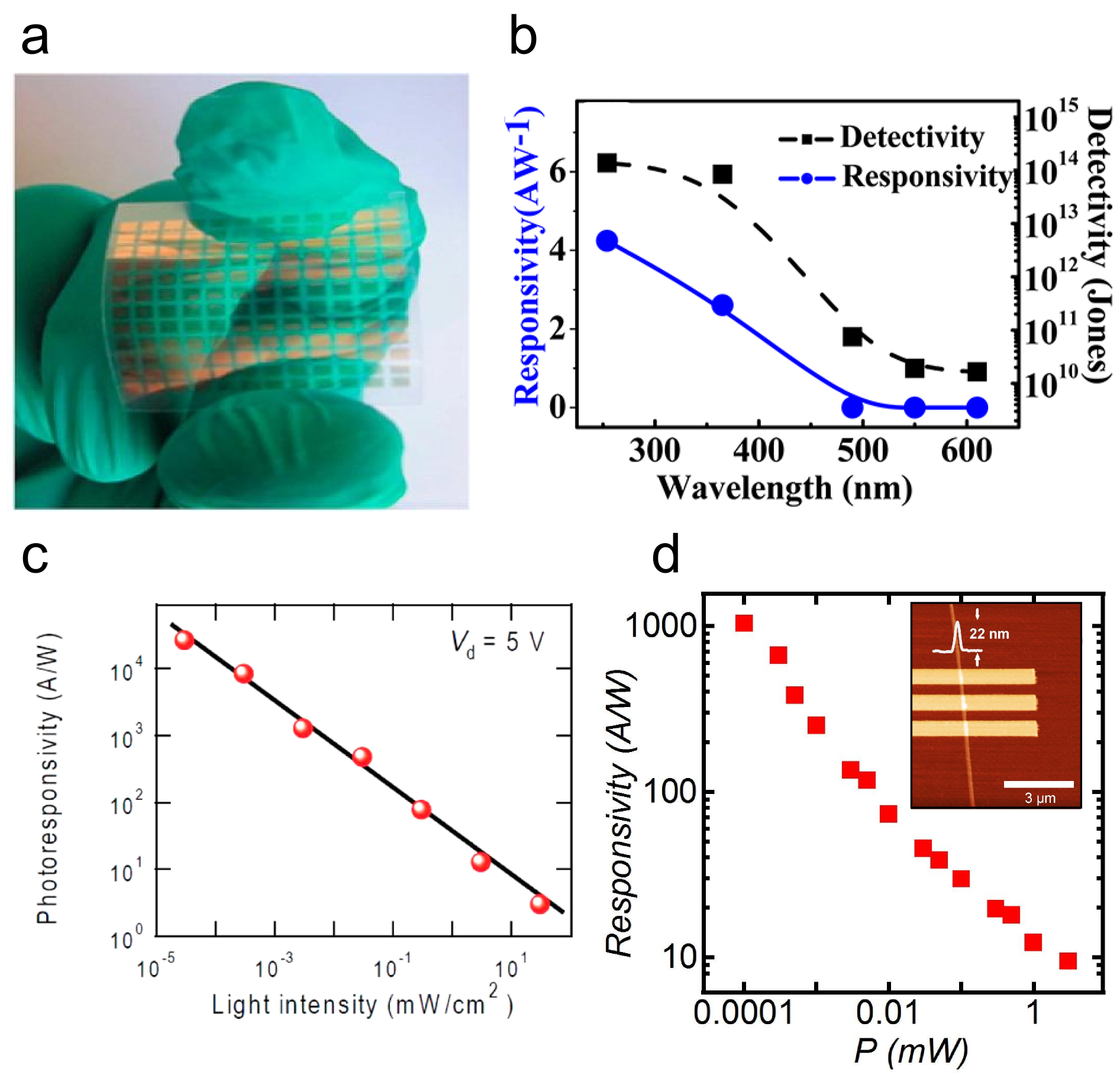}
\caption{(a) Photograph of a flexible photodetector based on \ce{GaS} flakes. (b) Responsivity (left axis) and Detectivity (right axis) of a \ce{GaS} photodetector as a function of excitation wavelength. (c) Log-log plot of the responsivity as a function of excitation power for a \ce{GaTe} photodetector. (d) Log-log plot of the responsivity as a function of excitation power ($\lambda =$ \SI{640}{\nm}) for the single \ce{TiS3} NR device shown in the inset. Panels a and b are adapted with permission from ref. \cite{Hu2013}, copyright 2013, American Chemical Society. Panel c is adapted with permission from ref. \cite{Liu2013}, copyright 2013, American Chemical Society. Panel d is adapted with permission from ref \cite{Island2014}, copyright 2014, John Wiley \& Sons. }
\label{tis3_gas}
\end{figure}

Photodetectors in the form of photoFET have been realized with multilayer flakes of these materials on regular \ce{SiO2}/\ce{Si} substrates and on flexible substrates (see Figure \ref{tis3_gas}a), showing similar performances. \ce{GaSe} and \ce{GaS}\cite{Hu2012,Hu2013} reach responsivities of \SI{4}{\A\per\W} and detectivities of \num{2e14} $\frac{\mathrm{cm}\sqrt{\mathrm{Hz}}}{\mathrm{W}}$, in the UV-blue range (Figure \ref{tis3_gas}b). On the other hand, direct bandgap materials (\ce{GaTe} and \ce{In2Se3})\cite{Liu2013, Jacobs-Gedrim2013} show responsivities of about \SI{1e4}{\A\per\W} at low optical excitation power (Figure \ref{tis3_gas}c). The photocurrent raises sublinearly as a function of excitation power, indicating that trap states play an important role in the photoconduction mechanism.\cite{Liu2013, Jacobs-Gedrim2013,Hu2012,Hu2013} The time response ranges from \si{\ms} to \si{\s}.

A particular case is represented by \ce{InSe}. It has a direct bandgap of about \SI{1.3}{\eV} (bulk) and reducing its thickness to below \SI{\sim 6}{\nm} leads to a transition to an indirect bandgap of higher energy.\cite{Mudd2013} The performance of photodetectors based on 10-nm-thick \ce{InSe} flakes have been studied independently by Lei \emph{et al.}\cite{Lei2014} and Tamalampudi \emph{et al.}\cite{Tamalampudi2014}. In their detailed study, Lei and colleagues\cite{Lei2014} report photocurrent spectra of 4-layer-thick \ce{InSe} under excitation wavelengths up to \SI{800}{\nm}. The tail of the photocurrent \emph{vs.} wavelength spectra is well fitted by a parabolic relation, strongly suggesting the indirect nature of the bandgap in ultrathin \ce{InSe}. The responsivity of these devices is about \SI{35}{\respo} and the response time is in the order of \SI{0.5}{\ms}. In their study, Tamalampudi and colleagues\cite{Tamalampudi2014} fabricate photoFETs based on \SI{\sim 12}{\nm} thick \ce{InSe} on both an oxidized \ce{Si} wafer and a bendable substrate. They measure gate tunable responsivities up to about \SI{160}{\A\per\W} and time response in the order \SI{4}{\s}. Both the larger responsivity and the much longer response time compared with the study of Lei \emph{et al.}\cite{Lei2014} indicate that long-lived trap states enhance the photoresponse in the devices studied by Tamalampudi \emph{et al.}\cite{Tamalampudi2014}

Very recently, also \ce{Sn} chalcogenide layered compounds have been used in photodetectors applications. Specifically, \ce{SnS2} has attracted recent attention as semiconductor material for transistors and photodetector. Exfoliated flakes have been recently employed as channel materials in FETs\cite{De2013} while chemically synthesized nanoparticle films of \ce{SnS2} have already shown promising photoresponse.\cite{Tao2014a} Su \emph{et al.}\cite{Su2014} have grown flakes of \ce{SnS2} via a seeded CVD technique and used them as channel material for photoFET. \ce{SnS2}-based photodetectors show gate-tunable responsivity up to about \SI{8}{\respo} and response time of \SI{\sim 5}{\us}, indicating a very fast photoresponse. Table \ref{novel_mat} summarizes the main figures-of-merit of detectors based \ce{Ga}, \ce{In} and \ce{Sn} chalcogenides.

In conclusion, \ce{Ga}, \ce{In} and \ce{Sn} compounds show responsivities that are comparable or larger than the one measured from TMDC based photoFETs. Moreover, their operation can be extended to the UV region of the spectrum, a feature that can enable them to be used in UV detectors. \ce{InSe} shows a direct-to-indirect transition for the bandgap when its thickness is reduced below \SI{6}{\nm}, resulting in a parabolic absorption tail up to \SI{1.4}{\eV} photon energy. Thicker \ce{InSe} flakes show large responsivity (about \SI{160}{\A\per\W}) and response time in the order of few seconds, indicating a strong effect of trap states on their photoresponse. Detectors based on \ce{SnS2} can reach responsivity of about \SI{8}{\respo} and fast response time, making them promising as fast detectors. 

\section{\ce{Ti}, \ce{Hf} and \ce{Zr} trichalcogenides photodetectors}\label{sec_tri}
In the previous Section, we reviewed the state-of-art in photodetection with layered semiconducting TMDCs and \ce{Ga}, \ce{In} and \ce{Sn} chalcogenide which demonstrate large responsivities in the visible range and have shown the possibility of realizing versatile devices through local gating. However, there are even more semiconducting chalcogenide compounds whose electronic properties in the ultra-thin limit are not yet explored. 

In this Section, we review the latest literature on photodetectors based on transition metal \emph{tri}chalcogenides materials: \ce{TiS3},\cite{Island2014} \ce{HfS3},\cite{Xiong2014} and \ce{ZrS3}.\cite{Tao2014} These trichalcogenide compounds present a direct bandgap with energy between \SI{1}{\eV} and \SI{3.1}{\eV}. The values are reported in Table \ref{bgap_tri_III}.  Island \emph{et al.}\cite{Island2014} have studied the photoresponse of ultrathin \ce{TiS3} nanoribbons (NR) and achieved responsivity up to \SI{2910}{\A\per\W} at \SI{640}{\nm} illumination wavelength, photoresponse to wavelengths up to \SI{940}{\nm} and short response times (fall time \SI{\sim 5}{\ms}). The devices were fabricated on isolated \ce{TiS3} nanoribbons via electron-beam lithography (see inset of Figure \ref{tis3_gas}d). 
As grown \ce{TiS3} ribbons (\num{100}\SI{\sim 300}{\nm} thick, \num{1}\SI{\sim 10}{\um} wide and \num{10}\SI{\sim 100}{\um} long) were mechanically exfoliated (to a thickness down to \SI{13}{\nm}) and transferred to a \SI{285}{\nm} \ce{SiO2}/\ce{Si} substrate.\cite{Island2014,Castellanos-Gomez2014} Figure \ref{tis3_gas}d shows a log-log plot of the responsivity as a function of excitation power for a representative device from reference \cite{Island2014}. The responsivity reaches about \SI{1000}{\A\per\W} at low power and decreases with increasing optical power. Again, the large responsivity and the decrease of the responsivity with optical power suggest that trap states play a large role in the photoreponse of ultrathin \ce{TiS3} NR.

\begin{table}

\centering

    \begin{tabular}{@{}cccc@{}}\toprule
     Material & Bandgap (\si{\eV})  & Remarks  & Ref \\
     \midrule  
    \ce{GaTe} & \tnrr{1.7} & direct & \cite{Liu2013} \\
    \ce{GaSe} & \tnrr{2.11} & direct at \SI{2.13}{\eV}& \cite{Hu2012} \\
    \ce{GaS} & \tnrr{2.59}  & direct at \SI{3.05}{\eV} & \cite{Hu2013} \\
    \ce{In2Se3} & \tnrr{1.3} & direct, $\alpha$-phase & \cite{Jacobs-Gedrim2013} \\
    \ce{InSe} & \tnrr{1.3}  & direct & \cite{Tamalampudi2014,Lei2014} \\
    \ce{SnS2} & \tnrr{2.2} & direct & \cite{Su2014,Tao2014a} \\
    \ce{TiS3} & \tnrr{1} &  direct & \cite{Island2014}  \\
    \ce{ZrS3} & \tnrr{2.56} & direct &\cite{Tao2014}  \\
    \ce{HfS3} & \tnrr{3.1} & direct &\cite{Schairer1973} \\
    
    \bottomrule
    \end{tabular}
\caption{Bulk bandgap of \ce{Ga}, \ce{In} and \ce{Sn} chalcogenides and \ce{Ti}, \ce{Hf} and \ce{Zr} trichalcogenides. }
\label{bgap_tri_III}
\end{table}

To obtain single \ce{HfS3} ribbons, Xiong \emph{et al.}\cite{Xiong2014} dispersed part of the as-grown material in ethanol and drop-cast it onto a \SI{300}{\nm} \ce{SiO2}/\ce{Si} substrate. Next, photolithography and metal evaporation were used to contact the randomly dispersed nanoribbons. The fabricated devices showed p-type behavior with ON currents in the order of few \si{\pico\A}. Under illumination, the current reached up to \SI{200}{\pico\A}; the responsivity is about \SI{110}{\respo} and the time response is about \SI{0.4}{\s}. The measured light absorbance extends up to about \SI{650}{\nm}.\cite{Xiong2014}

Tao \emph{et al.}\cite{Tao2014} studied the photoresponse of a network of \ce{ZrS3} ribbons transferred onto flexible substrates, polypropylene and paper. Electrical devices were fabricated by shadow-mask evaporation and, under illumination, demonstrate a responsivity of about \SI{5e-2}{\respo}, a photoresponse to excitation wavelengths up to \SI{850}{\nm} and a time response of \SI{13}{\s}.\cite{Tao2014}

In summary, from the available data, we conclude that photodetectors based on a single nanoribbon of a trichalcogenide compound can achieve responsivity comparable or larger than single layer chalcogenides ($R_\mathrm{MoS_2}$\ \SI{\sim 1000}{\A\per\W}) and faster response times ($\tau_\mathrm{MoS_2}$\ \SI{\sim 50}{\ms}). Table \ref{materials_tric_table} summarizes the main figures of merit of photodetectors based on these materials. The large responsivity and fast response make \ce{TiS3} a material promising for nanoscale photodetection, highlighting the need for further studies.

\section{Few-layer black phosphorus}\label{bP_photodetectors}

The recently re-discovered few-layer black phosphorus (bP) is an interesting material for photodetection especially due to its intermediate bandgap between graphene and TMDCs; its reduced bandgap compared to TMDCs make few-layer bP a promising candidate to extend the detection range with sizable responsivity that is achievable with 2D materials. Black phosphorus is an elemental layered compound composed of phosphorus atoms arranged in a puckered unit cell. 
Its bulk form has been studied in the 80's\cite{Morita1986} and, electronically, it behaves as a p-type semiconductor with mobilities\cite{Akahama1983} in the order of \SI{10000}{\mobility} and a bandgap of about \SI{0.35}{\eV}. Once exfoliated to thin layers, it shows ambipolar transport \cite{Castellanos-Gomez2014a,Buscema2014,Buscema2014b, Xia2014,Liu2014,Qiao2014,Li2014,Koenig2014}, mobilities\cite{Li2014,Xia2014} up to \SI{1000}{\mobility} and absorption edge around \SI{0.3}{\eV} in agreement with its bulk bandgap.\cite{Xia2014} The reduced symmetry of the puckered layer structure gives rise to strong anisotropy in the properties of few-layer bP, as evidenced by the anisotropic mobility and optical absorption. \cite{Xia2014}

Calculations predict that the bandgap of bP depends strongly on the number of layers and it should reach more than \SI{1}{\eV} once exfoliated down to a single layer.\cite{Castellanos-Gomez2014a,Liu2014, Tran2014a} 
Black phosphorus can, thus, bridge the gap between the large-bandgap TMDCs and zero-bandgap graphene, completing the high-responsivity detection range that is achievable with 2D layered materials. 

\paragraph*{Few-layer bP photodetectors.~~}
The research effort in bP-based phototodetectors has been extensive in the past year.\cite{Buscema2014,Buscema2014b,Engel2014a,Youngblood2015} Phototransistors based on ultrathin (\SIrange{3}{8}{\nm} thick) bP have been fabricated on \ce{SiO2}/\ce{Si} substrates. In the dark, bP-based photo-FETs readily achieve ambipolar transport via back-gating with hole mobilities in the order of \SI{100}{\mobility} and ON/OFF ratio \num{>1e3}.\cite{Buscema2014} The authors in this study\cite{Buscema2014} report on the performance of the bP photo-FET as a function of excitation wavelength, power and frequency. The responsivity reaches \SI{4.8}{\respo}, the photodetector wavelength range extends up to excitation wavelengths of \SI{940}{\nm} and the rise and fall times are \num{1} and \SI{4}{\ms}, respectively.\cite{Buscema2014} The readily achieved ambipolarity, the sizable responsivity across a wide wavelength range and the fast response make few-layer bP a promising material for photodetection applications up to the NIR part of the electromagnetic spectrum.

\begin{figure}[h!]

\captionsetup{width=.9\textwidth}
\centering
\includegraphics[width= \textwidth]{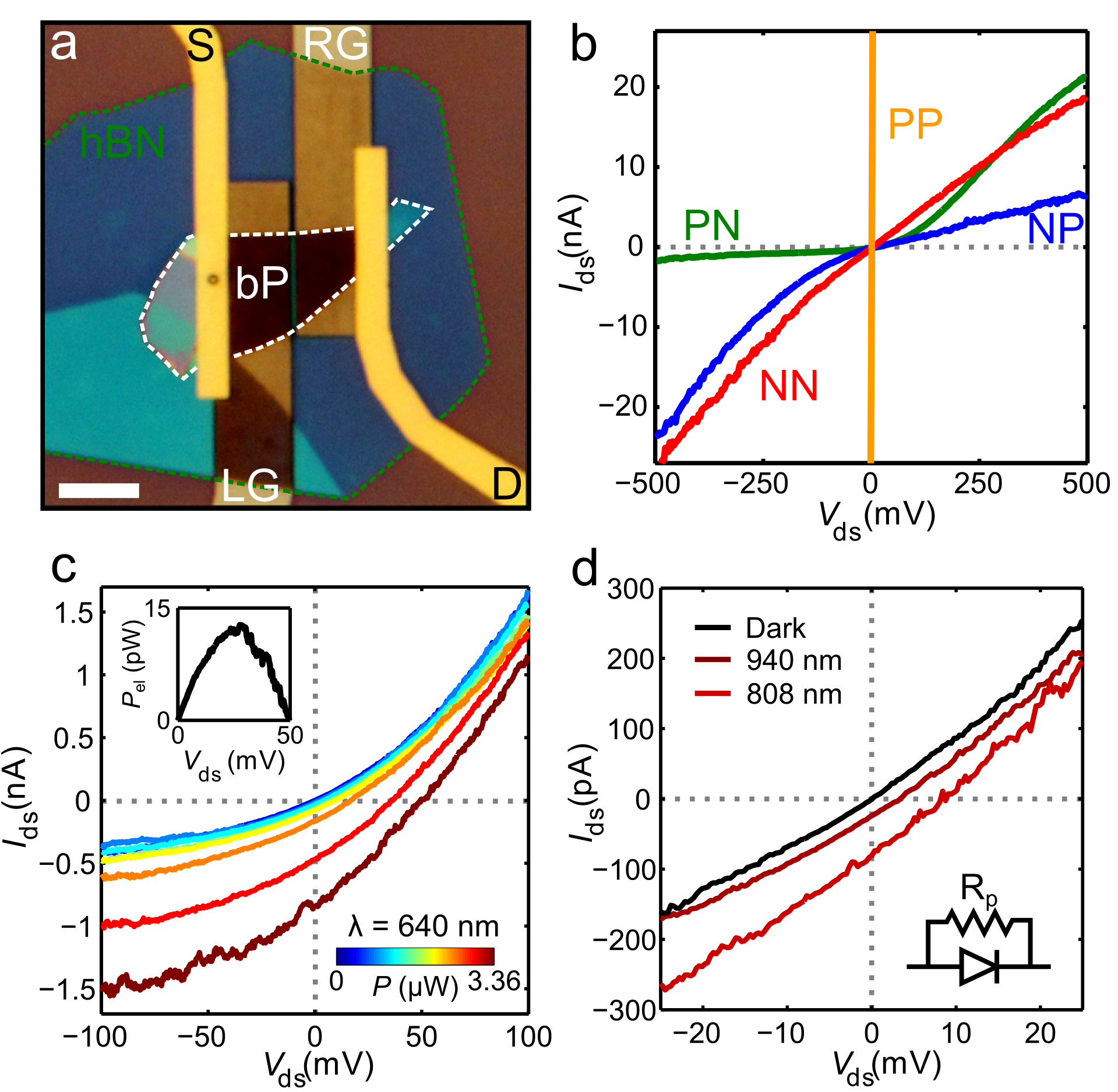}
\caption{(a) Optical image of one of the fabricated devices. (b) \ids$-$\vds\ characteristics of the device in different gate configurations. (c) \ids$-$\vds\ characteristics of the device in PN configuration under illumination with $\lambda$ \SI{=640}{\nm} and variable power. Inset: electric power generated by the device (d) Zoomed-in \ids$-$\vds\ characteristics of the device in PN configuration in dark (solid black line), under illumination with $\lambda$ \SI{=940}{\nm} (solid dark red line) and $\lambda$ \SI{=808}{\nm} (solid light red line). Inset: equivalent circuit, evidencing the resistance in parallel to the junction $R_\mathrm{p}$. All panels are adapted by permission from Macmillan Publishers LTD: Nature Communications, ref. \cite{Buscema2014b}, copyright 2014. }
\label{summary_bp1}
\end{figure}

\paragraph*{Electrostatically defined PN junction in few-layer bP.~~} Building on the ambipolarity, Buscema and Groenendijk \emph{et al.}\cite{Buscema2014b} have developed a PN junction via local electrostatic gating. The devices, based on two local split gates, combine the properties of two different 2D materials: hBN as an atomically-flat and disorder-free gate dielectric and few-layer bP as an ambipolar channel material. The devices have been fabricated by exploiting the deterministic transfer method.\cite{Castellanos-Gomez2014} Figure \ref{summary_bp1}a shows one of the fabricated devices.

Local electrostatic gating allows effective control over the type and concentration of charge carriers, enabling a versatile electrical behavior. Figure \ref{summary_bp1}b plots the \ids$-$\vds\ characteristics of the final locally-gated device in different gate configurations. With the two gates biased in the same polarity (PP or NN configuration), the device shows metallic behavior with linear \ids$-$\vds\ curves. When the two gates are set at opposite polarities (PN or NP configuration), the \ids$-$\vds\ curves become rectifying with the direction of the current controlled by the gate polarities.

Under increasing illumination intensity, the \ids$-$\vds\ curves display increasing $\isc$ and reverse current, as shown in Figure \ref{summary_bp1}c. This is consistent with a photovoltaic mechanism where the generated photocurrent adds to the reverse-bias current. Thus, an extra forward bias is needed to suppress the current flow, giving rise to a non-zero $\voc$. The inset of Figure \ref{summary_bp1}c shows the electrical power ($P_{\mathrm{el}} = V_{\mathrm{ds}} \cdot I_{\mathrm{ds}}$) generated by the device, which reaches about \SI{13}{\pico\W} under the largest illumination power.

Near-infrared (NIR) photons give also rise to photocurrent in the bP PN junctions. The \ids$-$\vds\ curves in the PN configuration in the dark and with excitation wavelengths of \SI{808}{\nm} and \SI{940}{\nm} ($P_\mathrm{device} =$ \SI{0.33}{\uW}) are presented in Figure \ref{summary_bp1}d.

Compared to PN junctions realized with \ce{WSe2},\cite{Pospischil2014,Baugher2014,Groenendijk2014} bP PN junctions present a more extended wavelength operation range, comparable $EQE$ but a lower $\voc$. The lower open-circuit voltage is consistent with the smaller bandgap of few-layer bP compared to that of \ce{WSe2}. Another limitation to the $\voc$ comes from the non-ideal diode behavior evidenced by the slope of the \ids$-$\vds\ characteristics at \vds\ \SI{=0}{\V}, both in the dark and under illumination. This indicates the presence of leakage paths for the generated carriers in parallel to the junction, which can be modelled by a parallel resistance ($R_\mathrm{p}$ in the inset of Figure \ref{summary_bp1}d). The value of $R_\mathrm{p}$ can be extracted by fitting the \ids$-$\vds\ characteristics with a modified Shockley model.\cite{Ortiz-Conde2000} For \ce{WSe2} PN junctions, $R_\mathrm{p}$ exceeds \SI{100}{\giga\ohm}\cite{Lopez-Sanchez2013,Pospischil2014,Baugher2014,Groenendijk2014} while for bP devices it is in the order of \SI{100}{\Mohm},\cite{Buscema2014b} evidencing the more ideal behaviour of the \ce{WSe2} devices.

\begin{figure}[h!]
\captionsetup{width=.9\textwidth}
\centering
\includegraphics[width= \textwidth]{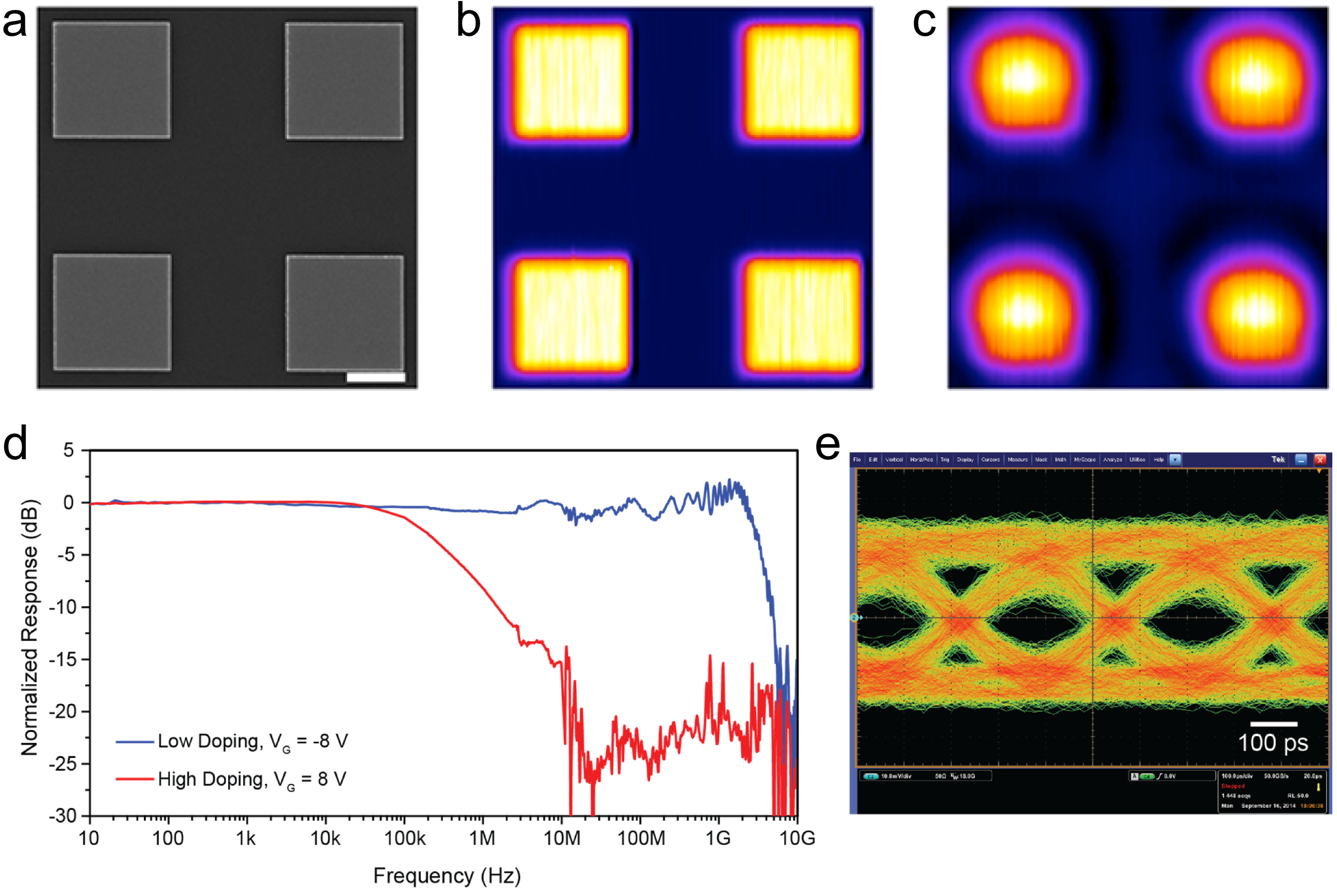}
\caption{(a) Scanning electron micrograph of the target image object. The scale bar is \SI{2}{\um}. (b) Image of the object in (a) obtained under focused illumination with $\lambda$\ \SI{=532}{\nm} and reading the reflected signal with a photodetector based on \SI{120}{\nm} thick bP. (c) Image of the object in (a) obtained under focused illumination with $\lambda$\ \SI{=1550}{\nm} and reading the reflected signal with a photodetector based on \SI{120}{\nm} thick bP. (d) Photoresponse as a function of modulation frequency at low doping (blue curve) and high doping (red curve). (e) Eye-diagram for 3 Gbits/s communication. Panels (a),(b),(c) are adapted with permission from ref. \cite{Engel2014a}, copyright 2014, American Chemical Society. Panels (d and e) are adapted by permission from Macmillan Publishers Ltd: Nature Photonics, ref. \cite{Youngblood2015}, copyright 2015.}
\label{materials_bp_2}
\end{figure}

\paragraph*{Photothermoelectric effect in few-layer bP.~~} The thermoelectric effect plays a major role in the photocurrent generation of ultrathin bP devices. Hong \emph{et al.}\cite{Hong2014} performed scanning photocurrent microscopy (SPCM) on a \SI{8}{\nm} high bP transistor with a diffraction limited $\lambda$\ \SI{=780}{\nm} laser spot as a function of polarization and gate bias. The authors find that both the PV effect (Schottky barriers) and the PTE effect contribute to the photocurrent generation.\cite{Hong2014} In the OFF state, the electric field at the Schottky barriers separates the photogenerated electron-hole pairs. Applying a gate bias brings the device in the ON state and drastically reduces the contact resistance, allowing the PTE effect to contribute the photocurrent. By employing the Mott formalism (Equation \ref{Mott2}) the authors estimate a maximum Seebeck coefficient of \SI{\sim 100}{\seebeck}, which is reduced by a factor 10 with increasing gate bias. Recently, the bulk Seebeck coefficient of bP was directly measured\cite{Flores2015} to be \SI{350}{\seebeck}, illustrating the importance of the PTE effect in bP.

In the OFF state, where the Schottky-Barrier induced PV is dominant, the photocurrent magnitude is modulated by incident light polarization. When the polarization is aligned with the low-effective-mass crystallographic axis (x-axis) the photocurrent is enhanced; conversely, aligning the polarization with the high-effective-mass axis (y-axis) suppresses the photocurrent. The polarization-dependent photocurrent is a direct result of the strongly asymmetric band structure of few-layer bP, highlighting a difference from the isotropic TMDCS.

\paragraph*{Imaging applications of bP photodetectors.~~}The fast development of photodetectors based on bP has led to the realization of high performance devices for imaging\cite{Engel2014a,Low2014} and high-speed communication\cite{Youngblood2015}.
Engel \emph{et al.}\cite{Engel2014a} have achieved high-resolution imaging with a phototransistor based on a \SI{120}{\nm} thick bP flake, working via the photo-bolometric effect.\cite{Engel2014a,Low2014} Figure \ref{materials_bp_2}a shows the geometry of the objects to be imaged: \SI{4}{\um} wide metallic squares deposited on glass and separated by \SI{2}{\um}. A focused laser spot is scanned over the surface of the object and the reflected signal is read by the bP phototransistor. Figure \ref{materials_bp_2}b,c show the images obtained by reading the electric signal from the bP photodetector at a wavelength of \SI{532}{\nm} and \SI{1550}{\nm}, respectively. The large contrast allows for clear imaging of the features. Feature visibility is constant for sizes down to \SI{1}{\um} with $\lambda$\ \SI{=532}{\nm} and decreases by less than 20\% with $\lambda$\ \SI{=1550}{\nm}.\cite{Engel2014a}

\paragraph*{High-speed applications of bP photodetectors.~~} The study by Youngblood \emph{et al.}\cite{Youngblood2015} shows the photoresponse for a \SI{\sim 11}{\nm} bP flake up to \si{\GHz} frequencies.\cite{Youngblood2015} The few-layer bP is embedded in an on-chip waveguide structure with a few-layer graphene top gate, allowing for optimal interaction with light and tunability of the carrier density. The DC responsivity reaches \SI{130}{\respo}. Figure \ref{materials_bp_2}d shows the photoresponse of the few-layer bP device as a function of modulation frequency for low doping (blue curve) and high doping (red curve). The response rolls off at about \SI{3}{\GHz} for low doping and at \SI{2}{\MHz} for high doping. The authors ascribe this difference to a change of photocurrent generation mechanism induced by the doping.\cite{Youngblood2015} Figure \ref{materials_bp_2}e shows an eye-diagram at 3 Gbits/s data rate (similar to the one reported for a graphene-based device in ref. \cite{Mueller2009}). The clean eye diagram indicates that the photodetector can operate well at this high data transfer rate. This study shows that few-layer bP phototransistors, embedded in engineered devices, can achieve data transfer speed similar to graphene photodetectors \cite{Mueller2010,Xia2009} with larger responsivities.

Table \ref{materials_bP_table} summarizes the principal figures of merit of bP based detectors.

\section{Device comparison}\label{discussion}
\captionsetup{width=.9\textwidth}
\begin{figure*}[t]
\centering
\includegraphics[width = .85\textwidth]{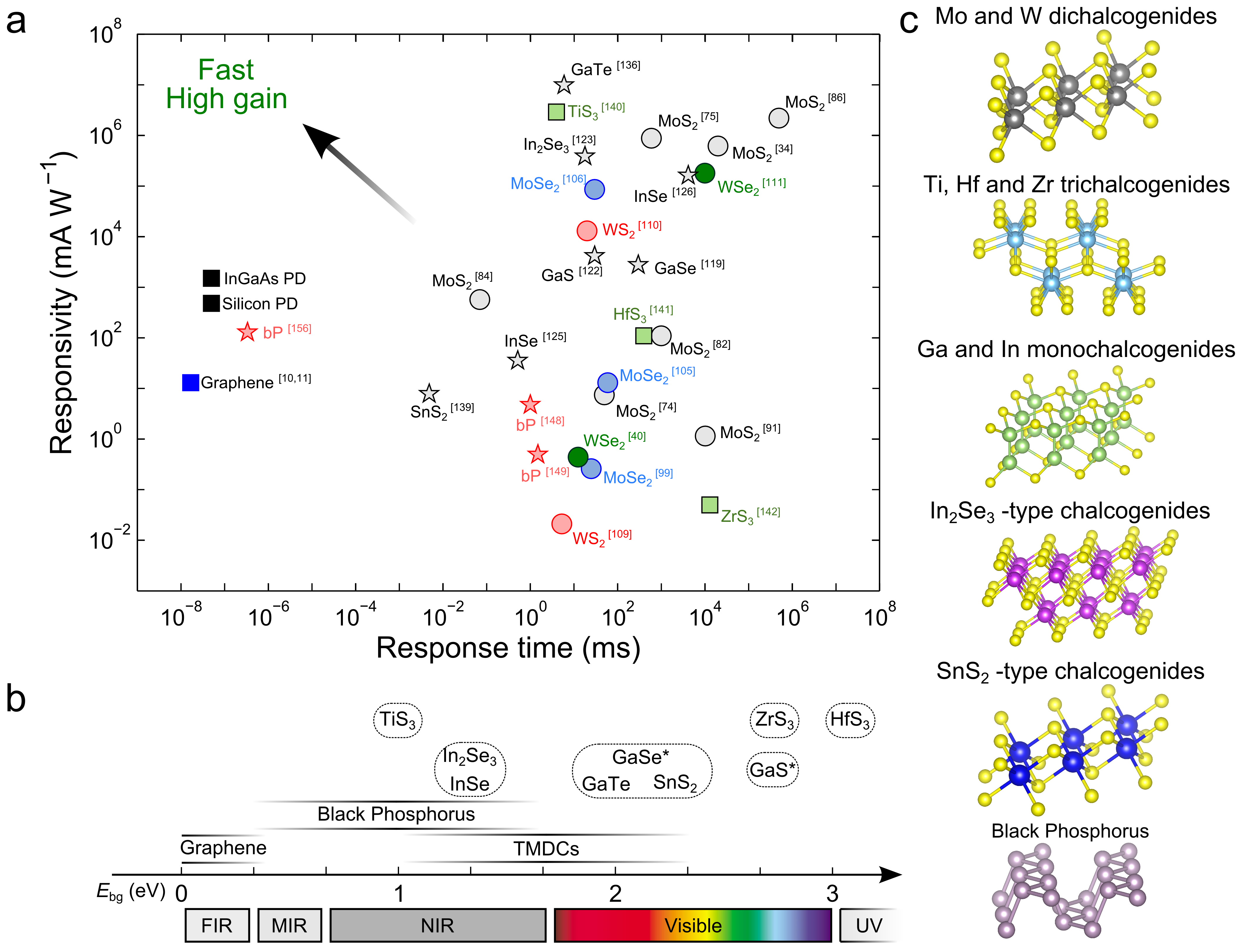}
\caption{(a) Responsivity against response time for the reviewed devices, for a commercial silicon and \ce{InGaAs} photodiodes (black squares) and for graphene photodetectors (blue square). (b) Bandgap of the different layered semiconductors and electromagnetic spectrum. The shaded lines above and below graphene, TMDCS and black phosphorus roughly indicate the range of bandgap values that can be addressed by the various families of layered materials. The exact bandgap value would depend on the number of layers, strain level and chemical doping. The asterisk indicates that the material's fundamental bandgap is indirect. FIR: far infrared; MIR: mid infrared; NIR: near infrared; UV: ultraviolet. (c) Overview of the crystal structures of the different families of layered materials. The yellow atoms always corresponds to a chalcogen species.}
\label{fig_dev_comparison}
\end{figure*}

In this Section, we summarize and compare the responsivity and response time of the devices we have so far reviewed. Table \ref{materials_mos2_table} summarizes the responsivity, rise time and spectral range for photodetectors based on \ce{MoS2}, \ce{MoSe2}, \ce{WS2} and \ce{WSe2}. Within devices based on the same material, there is a large variability in the responsivity and rise time figures-of-merit, likely indicating that the device fabrication, measurement conditions and contact metals have a large effect on the photoresponse. It is worth mentioning that, in general, most of the photodetectors based on \ce{MoS2} present orders-of-magnitude larger responsivity and response times with respect to photodetectors based on \ce{MoSe2} or \ce{WS2}, possibly indicating that trap states in \ce{MoS2} are longer-lived than in the other compounds. All these compounds show a remarkable effect of the environment in which the measurements are performed, suggesting their potential application as light-sensitive gas detectors.

PN-junctions based on single-layer \ce{WSe2} are also included in Table \ref{materials_mos2_table} even though their working principle is based on the photovoltaic effect, rather than photoconduction/photogating. Their responsivity is in the order of \num{1} to \SI{10}{\respo} when operated in the photovoltaic mode and shows a small increase in the photoconductive mode. The reported responsivity for the \ce{WSe2} PN junctions (see Table \ref{materials_mos2_table}) is generally lower than photodetectors based on other TMDCs, especially in comparison with \ce{MoS2}, indicating that the internal gain mechanism (trap states) is less efficient. This is expected for photodetectors working with the photovoltaic principles. However, locally-gated \ce{WSe2} devices show versatile behavior, combining a photo-FET with a gate-controlled photodiode in a single device.

Table \ref{materials_tric_table} summarizes the figures-of-merit for photodetectors based on  and \ce{Ti}, \ce{Hf} and \ce{Zr} trichalcogenides. Devices based on \ce{TiS3} ultrathin nanoribbons achieve both slightly higher responsivity, faster response and wider detection range than detectors based on \ce{MoS2}. This good performance, coupled to the reduced dimensionality, make \ce{TiS3} nanoribbons a promising material for nanostructured photodetection applications. On the other hand, both \ce{ZrS3}- and \ce{HfS3}-based detectors show poor responsivity and a slow response. 

Devices based on multilayers of \ce{Ga}, \ce{In} and \ce{Sn} chalcogenides show responsivities from tens to thousands \si{\A\per\W} and response time between few and hundred \si{ms}. For example, \ce{GaTe} photodetectors outperform detectors based on single-layer \ce{MoS2} reaching responsivities of \SI{1e4}{\A\per\W} and response time as short as \SI{6}{\ms}. Photodetectors based on \ce{In2Se3}, \ce{GaSe} and \ce{GaS} show responsivities and response times that are slightly better than most detectors based on \ce{MoS2}. Better control over device geometry, fabrication recipes and measurement conditions are needed for a proper benchmarking, but these materials have already demonstrated their promise for visible and UV detection. 

Both the semiconducting di- and tri-chalcogenides do not show photoresponse at telecommunication wavelengths. On the other hand, the newly (re)discovered black phosphorus demonstrates sizable responsivity (about \SI{0.1}{\A\per\W}) and response speed ($f_\mathrm{3dB}$ \SI{\sim 3}{\GHz}) under $\lambda$\ \SI{=1550}{\nm} excitation (see Table \ref{materials_bP_table}). Moreover, few-layer bP has shown ambipolar transport and the possibility of realizing versatile locally-gated devices, like in the case of single-layer \ce{WSe2}. This makes few-layer bP a promising candidate for fast and broadband detection and light energy harvesting in the IR part of the spectrum.
Compounds from elements in the III-VI groups (\ce{Ga}, \ce{In} with \ce{S}, \ce{Se}) are also attracting attention for photodetection applications. 
\begin{figure*}
\captionsetup{width=.9\textwidth}
\centering
\includegraphics[width= \textwidth]{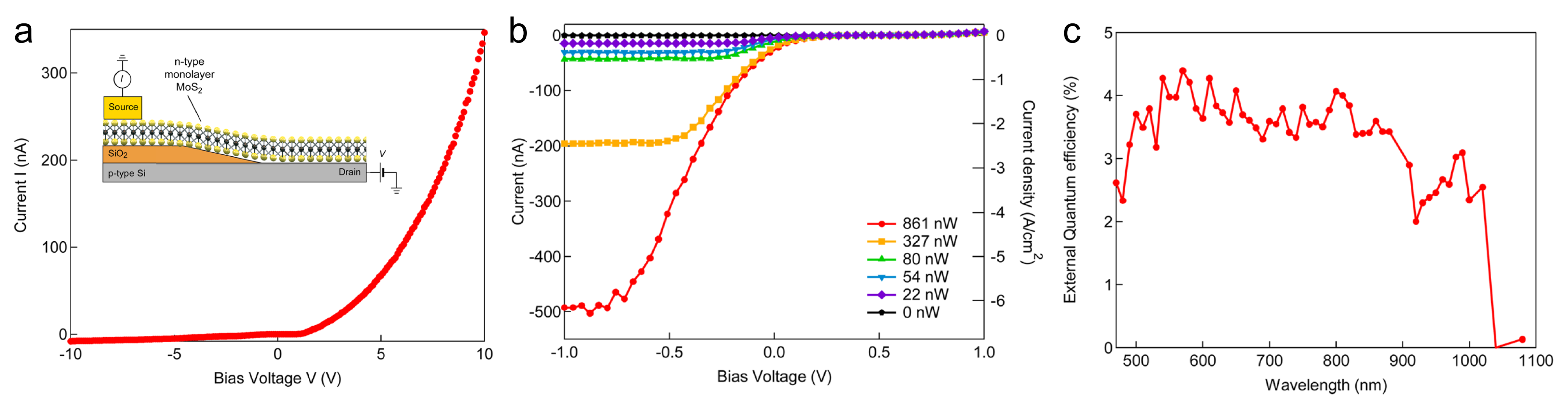}
\caption{(a) \ids$-$\vds\ characteristic in the dark showing diode-like behavior. Inset: Device schematics. (b) \ids$-$\vds\ under illumination with $\lambda$\ \SI{=541}{\nm} and varying optical power, indicated in the panel. (c) External quantum efficiency as a function of excitation wavelength. All panels are adapted from with permission from ref. \cite{Lopez-Sanchez2014}, copyright 2014, American Chemical Society. }
\label{mos2_si_hetero}
\end{figure*}

To give an overview of the different materials for photodetection, the data from Tables 4 to 7 is summarized in Figure \ref{fig_dev_comparison}a. As a benchmark, the responsivity and response time of a commercial silicon photodiode (black square) and a graphene photodetector (blue square) are shown as well. It is evident that photodetectors based on semiconducting layered materials display a large (about 10 orders of magnitude) variation in their responsivity. Regarding the response time, it appears that all but one of the reviewed devices show response times larger than \SI{\sim 1e-2}{\ms}. The rather long response times can be attributed to the presence of long-lived trap states (in devices with responsivity above \SI{1e3}{\respo}) or to the limitation of the measurement electronics (RC time) likely induced by the large resistance of the device or input impedance of the current-to-voltage amplifier used in the read-out electronics. 

The bP-based device presented in ref. \cite{Youngblood2015} stands out from the rest by showing comparable performances to both a commercial silicon photodiode and a graphene-based detector. In particular, this device shows responsivity and response times (at telecommunications wavelengths) that are within one order of magnitude from a commercial silicon photodiode (in the visible), strongly indicating that bP-based photodetectors can already compete with traditional silicon detectors.

In terms of absolute responsivity, roughly, half of the reviewed devices present values equal or larger than the \ce{Si} photodiode. Especially, devices based on \ce{MoS2}, \ce{WSe2}, \ce{InSe}, \ce{In2Se3}, \ce{GaTe} and \ce{TiS3} present a responsivity which is three to four orders of magnitude above. The large responsivity supports the claim that layered semiconducting materials hold promise for ultrasensitive applications in the visible, for which the response times in the order of \si{\ms} may be acceptable.

Figure \ref{fig_dev_comparison}b relates the bandgap energy of layered materials to the electromagnetic spectrum, showing that 2D semiconductors cover a very broad portion of the spectrum (from mid-infrared to the UV). We note that the exact bandgap value depends on the number of layers in a quantized fashion. To achieve continuous bandgap tunability one can employ strain engineering, chemical doping and (especially for the TMDCs) alloys of different materials, as recently proven for an alloy of \ce{MoSe2} and \ce{MoS2} by Li \emph{et al.}\cite{Li2014g}

 It is also shows that there are many materials with bandgap around \SI{1.3}{\eV}, a region of interest for photovoltaic applications. 

 Figure \ref{fig_dev_comparison}c gives an overview of the crystal structures for different families of layered materials.

\section{Future directions}

So far, we have reviewed photodetectors that are based on a single semiconducting material, used as channel material in a photo-transistor or a locally-gated PN junction. In this Section we will start by discussing one of the promising directions of photodetection with semiconducting layered materials: artificial heterostructures of different layered compounds. We will then continue by briefly summarizing the recent progress in other possible future directions: Surface decoration, light-matter interaction enhancement and suspended devices.

\subsection{Artificial van der Waals heterostructures.}
Deterministic transfer techniques\cite{Dean2010,Schneider2010,Song2011a,Zomer2011,Bonaccorso2012,Wang2013} have opened the door to the realization of artificial heterostructures based on single- or few-layer semiconducting materials. These heterostructures can be built on conventional (3D) materials (like \ce{Si}\cite{Lopez-Sanchez2014,Lopez-Sanchez2014a,Li2014e} and \ce{InAs}\cite{Chuang2013}) or on other two dimensional materials and have triggered a great deal of experimental work on their  electrical\cite{Britnell2013,Yu2013,Roy2013,Zhang2014b,Lee2014,Lee2014a,Furchi2014a,Fang2014,Huo2014a,Gong2014,Deng2014,Wi2014}, optical\cite{Lui2014,Fang2014a,He2014,Yu2015,Hong2014a,Tongay2014,Ceballos2014} and mechanical properties.\cite{Liu2014a} In this Section, we review the recent experimental work aimed to establish the opto-electronic performances of devices based on these van der Waals heterostructures.
\paragraph*{\ce{MoS2}/\ce{Si} heterostructures.~~}
A device based on single-layer \ce{MoS2} on p-doped \ce{Si} has been fabricated by Lopez-Sanchez \emph{et al.}\cite{Lopez-Sanchez2014,Lopez-Sanchez2014a} and Li \emph{et al.}\cite{Li2014e}. The inset of Figure \ref{mos2_si_hetero}a shows a typical device geometry: single-layer \ce{MoS2} is transferred on top of a highly p-doped \ce{Si} substrate and charge transport occurs across the \ce{MoS2}/\ce{Si} interface. In the dark, the resulting \ids-\vds\ characteristics are diode-like (Figure \ref{mos2_si_hetero}a), evidencing the formation of a PN junction between the n-type \ce{MoS2} and the p-doped \ce{Si}.

Under illumination, the \ids$-$\vds\ curves show an increased reverse-bias current and $\isc$ (Figure \ref{mos2_si_hetero}b), strongly suggesting that the photogenerated carriers are separated by the electric field generated at the interface between the single-layer \ce{MoS2} and the \ce{Si} surface. The EQE of this device reaches about 4\% in the visible and decreases to about 2.5\% at $\lambda$\ \SI{=1000}{\nm} (see Figure \ref{mos2_si_hetero}c).\cite{Lopez-Sanchez2014} Compared with detectors based on only single-layer \ce{MoS2}, the EQE is much lower; however, the wavelength detection range is extended due to the lower bandgap of silicon, enabling detection of NIR photons.

To conclude, devices based on \ce{MoS2}/\ce{Si} heterostructures show diode-like electrical characteristics, photovoltaic effect and electroluminescence, effects that strongly indicate an efficient charge transfer at the interface between the \ce{MoS2} and the \ce{Si} substrate. Moreover, they clearly proof the possibility to integrate layered semiconductors with current silicon technology.
\begin{figure*}[t]
\captionsetup{width=.9\textwidth}
\centering
\includegraphics[width= \textwidth]{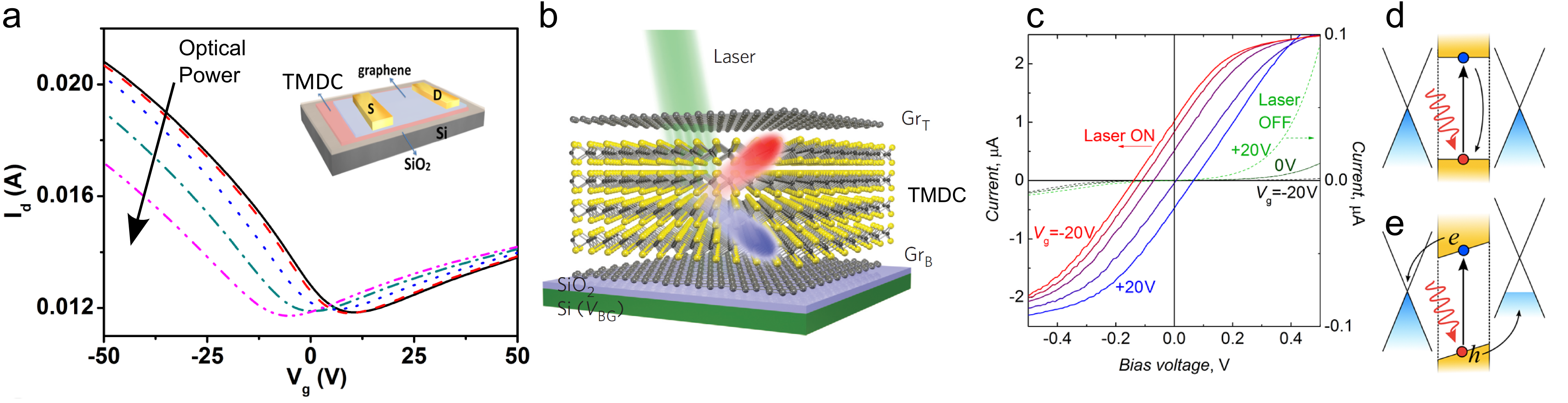}
\caption{(a) \ids$-$\vg\ characteristics for the device in panel (a) in the dark (solid black line) and under varying illumination intensities ($\lambda$ \SI{=650}{\nm}). Inset: device schematics for graphene/TMDC heterostructures. Note that charge transport is measured across the graphene layer. (b) Device schematics for graphene/TMDC/graphene heterostructure. Note that electrical transport is measured across the two graphene layers, across the TMDC. (c) \ids$-$\vds\ characteristics for device in panel (b) as a function of gate. The right (left) y-axis gives the current magnitude for measurements in dark, green traces (under illumination, blue to red traces). (d) Schematic band alignment in equilibrium. (e) Schematic band alignment under gate bias or chemical doping. Panels (a) is adapted by permission from Macmillan Publishers LTD: Scientific Report ref. \cite{Zhang2014b}, copyright 2014. Panel (b) is adapted by permission from Macmillan Publishers LTD: Nature Nanotechnology, ref. \cite{Yu2013}, copyright 2013. Panels (c), (d) and (e) are adapted ref. \cite{Britnell2013} with permission from AAAS. }
\label{graph_tmdc_hetero}
\end{figure*}

\paragraph*{Graphene/TMDCs heterostructures.~~}

The inset of Figure \ref{graph_tmdc_hetero}a shows a schematic of a typical graphene/TMDC heterostructure; a layer of a semiconducting TMDC is covered with single-layer graphene which is contacted by metallic electrodes. In this geometry, charge is transported in the graphene layer and the TMDC serves as an additional gate dielectric with light-sensitizing properties. With this geometry, one layer of graphene over a multilayer \ce{MoS2} resulted in photodetectors with drastically enhanced responsivity (\SI{\sim 5e8}{\A\per\W}) at the expense of an extremely slow response times (\SI{> 1e3}{\s}).\cite{Roy2013} The large responsivity stems from a strong photogating effect on the single-layer graphene channel induced by the localized states in the \ce{MoS2} layer. Under illumination, the photogenerated holes are trapped in localized states in the \ce{MoS2} while the photogenerated electrons are transferred to the graphene, free to circulate. The trapped holes reside longer in the \ce{MoS2} flake and thus act as a local gate on the graphene channel. Their long lifetime is responsible for the extremely large responsivity and long response times.

Photogating has also been observed in a hybrid detector composed of a single-layer \ce{MoS2} covered by a single layer graphene, both grown by CVD techniques.\cite{Zhang2014b} Under illumination, the \ids$-$\vg\ characteristics shift horizontally, clearly indicating that the photoresponse mechanism is dominated by the photogating effect (Figure \ref{graph_tmdc_hetero}a). Interestingly, the authors report a decrease of the magnitude of the photogating effect when the measurements are performed in vacuum. The smaller shift in vacuum is attributed to a reduction in the efficiency of the charge transfer between the \ce{MoS2} and the graphene. In air, the graphene is slightly more p-type, increasing the electric field at the interface and, hence, increasing the charge transfer rate for electrons. In vacuum, the graphene is slightly more n-type, suppressing the electron transfer, thus reducing the generated photocurrent.\cite{Zhang2014b} For this device, the responsivity is also very high (about \SI{1e7}{\A\per\W}) at low illumination power and response times range from tens to hundreds of seconds (both values for the device in air).

\paragraph*{Graphene/TMDCs/Graphene heterostructures.~~} In the previous Section, we have discussed two examples of graphene/TMDC heterostructures where the charge transport was measured in the graphene layer and the TMDC flake served as sensitizing material to enhance the photoresponse. In this Section, we discuss devices based on vertical stacks of graphene/TMDCs/graphene where charge transport occurs through the TMDC layer in the out-of-plane direction. In other words, the graphene layers are used as transparent contacts to the photoactive TMDC layer.

In similar studies, Britnell \emph{et al.}\cite{Britnell2013} and Yu \emph{et al.}\cite{Yu2013} fabricated vertical stacks of graphene, a \SI{\sim 50}{\nm} thick TMDC and another graphene layer on various substrates, including \ce{SiO2}, indium-tin-oxide (ITO) and poly-(ethylene terephthalate) (PET, a flexible and transparent polymer). Figure \ref{graph_tmdc_hetero}b shows a schematic of a typical device. In the dark, such devices behave as tunnel transistors in which the current can be controlled by the gate electric field (see Figure \ref{graph_tmdc_hetero}c, green traces).\cite{Britnell2013,Yu2013} Under illumination, the \ids$-$\vds\ characteristics become linear (until saturation) and show a gate-tunable $\isc$ indicating that the photovoltaic effect dominates the photocurrent generation (see Figure \ref{graph_tmdc_hetero}c, blue to red traces).\cite{Britnell2013,Yu2013}

The emergence of the photovoltaic effect can be understood by looking at the band diagram schematics presented in Figure \ref{graph_tmdc_hetero}d and Figure \ref{graph_tmdc_hetero}e. In an ideal case (Figure \ref{graph_tmdc_hetero}d) the band alignment between the graphene layers and the TMDC is symmetric. This symmetric alignment gives no preferential direction for the separation of the photogenerated electron-hole pairs, therefore leading to zero net photocurrent. On the other hand, a small difference in the doping level of the two graphene electrodes (induced by gating or chemical doping) will result a misalignment between the bands and, thus, in a built-in electric field. This built-in electric field separates the e-h pairs and generates the $\isc$ (see Figure \ref{graph_tmdc_hetero}e). As a consequence, the sign and magnitude of $\isc$ are gate-tunable, as evidenced from the blue-to-red traces in Figure \ref{graph_tmdc_hetero}c. In both studies, SPCM measurements reveal that the photocurrent is generated only in the regions of the heterostructure where such an asymmetry is realized. 

The responsivity of these devices is in the order of \SI{0.2}{\A\per\W} and the maximum electrical power that can be generated is in the order of a few \si{\micro\W}, much larger than current PN junctions based on single-layer \ce{WSe2}\cite{Baugher2014,Pospischil2014, Groenendijk2014}, making this type of devices very promising candidates for flexible solar cells.

\paragraph*{bP/TMDC heterostructures.~~}
Also heterostructures comprising few-layer bP (p-type) and single layer \ce{MoS2} (n-type) have recently shown gate-tunable photovoltaic effect, generating about \SI{2}{\nano\W} of electrical power.\cite{Deng2014}

\paragraph*{TMDC/TMDC heterostructures.~~}
\begin{figure}[h!]
\captionsetup{width=.9\textwidth}
\centering
\includegraphics[width= \textwidth]{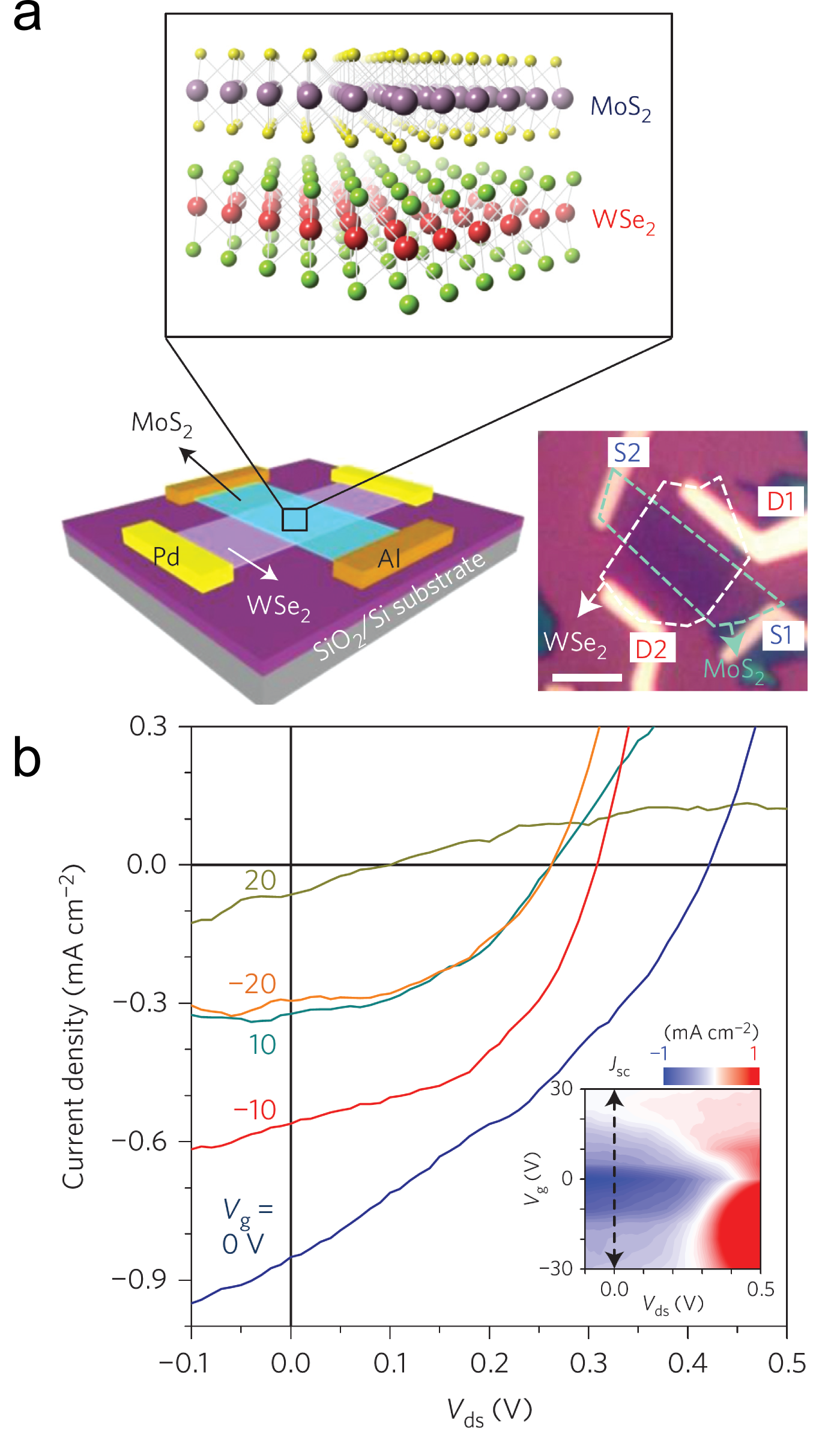}
\caption{(a) Device schematic. Top: schematic representation of the single layers composing the heterostructure. Left: schematic of the contact geometry. \ce{Pd} (\ce{Al}) is used to contact the single-layer \ce{WSe2} (\ce{MoS2}). Right: optical micrograph of the final device. D1 and D2 (S1 and S2) indicate the contacts to the \ce{WSe2} (\ce{MoS2}). The scale bar is \SI{2}{\um}. (b) \ids$-$\vds characteristics under illumination at different gate voltages. Inset: false color map of the current under illumination as a function of \vds\ and \vg. Panels (a) and (b) are adapted by permission from Macmillan Publishers LTD: Nature Nanotechnology, from ref. \cite{Lee2014a}, coypright 2014.}
\label{tmdcs_hetero}
\end{figure}
Very recently, heterostructures based on single-layered semiconducting TMDCs have also been studied. For example, heterostructures of single-layer \ce{MoS2} and single-layer \ce{WSe2} have been fabricated by several groups.\cite{Lee2014,Lee2014a,Furchi2014a,Fang2014} Figure \ref{tmdcs_hetero}a shows an example of such a heterostructure from the Columbia group.\cite{Lee2014a} The \ce{WSe2} and \ce{MoS2} single layers have been chosen because of their natural tendency to display p- and n-type conduction, respectively. In Figure \ref{tmdcs_hetero}a, single-layer \ce{MoS2} and single-layer \ce{WSe2} are crossed on top of each other and are contacted with \ce{Al} and \ce{Pd} to enhance the n-type and p-type character, respectively

In dark, \ce{WSe2}/\ce{MoS2} heterostructures show gate-tunable rectifying behavior.\cite{Lee2014a,Furchi2014a} Under illumination, these heterostructures show a gate-dependent photovoltaic effect as evidenced by the \ids$-$\vds\ characteristics in Figure \ref{tmdcs_hetero}b. The $\isc$ and $\voc$ reach a maximum at \vg\ \SI{=0}{\V}, the position at which the rectifying ratio is the highest.\cite{Lee2014a,Furchi2014a} The inset of Figure \ref{tmdcs_hetero}b shows a colormap of the measured current as a function of \vg\ and \vds, highlighting the reduction of $\isc$ as the magnitude of the gate voltage increases. Through spatial photocurrent and photoluminescence maps the authors ascribe the origin of this photovoltaic effect to charge transfer between the \ce{WSe2} and \ce{MoS2}.\cite{Lee2014a,Furchi2014a,Fang2014}

Huo \emph{et al.}\cite{Huo2014a} have recently fabricated heterostructures based on multilayer \ce{MoS2} and \ce{WS2} and measured the same qualitative behavior.\cite{Huo2014a} CVD techniques have also been used to grow both vertical and lateral heterostructures.\cite{Gong2014} CVD-grown devices showed stronger interlayer interaction compared with their transferred counterparts,\cite{Gong2014} likely being caused by an improved charge transfer. An increase in the charge-transfer efficiency in CVD-grown \ce{MoS2}/\ce{WSe2} heterostructure will likely lead to a larger rectification and stronger photovoltaic effect. In conclusion, TMDC/TMDC heterostructures present gate-tunable and strong photovoltaic power generation (\SI{\sim 6}{\micro\W}) coupled to the possibility of direct CVD growth; these properties make them hold promise for large-area flexible and transparent solar cells.

\paragraph*{CVD techniques.~~}

In recent years, there has been an intense effort in studying the CVD growth of layered compounds, especially graphene, hexagonal boron nitride and transition metal dichalcogenides.
CVD growth of the semiconducting TMDCs has proven to be a reliable technique that delivers high quality large-area single layers as shown in ref \cite{Zande2013}. Van der Zande's seminal work demonstrates that, with a straightforward recipe, it is possible to obtain isolated triangular single-layer \ce{MoS2} flakes with edges length in the order of \SI{100}{\um} at the side of a continuous single-layer film. The quality of the as-grown single layers is proven via luminescence and transmission electron microscopy. The other Mo- and W- based dichalcogenides can also be grown via CVD techniques $-$ See refs. \cite{Zhang2013c,Shaw2014,Cong2014, Zhou2015, Huang2015}. For more extensive details on the CVD growth of single-layers of TMDCs and other layered semiconductors we refer the reader to very recent in-depth reviews.\cite{Shi2015,Ji2015}. The tables presented in ref. \cite{Shi2015} give a comprehensive and clear overview of the growth conditions and properties of the resulting material.

These high quality CVD-grown single layers are then used to fabricate large-area heterostructures via deterministic transfer methods. Figure \ref{new_future}a shows such a vertical heterostructure composed by CVD grown \ce{MoS2} and \ce{WSe2} single layers built on sapphire. The overlap area between the flakes is several square microns large.\cite{Chiu2014} These heterostructures have been mostly employed in luminescence/Raman studies,\cite{Hong2014a,Zande2014,Chiu2014} second harmonic generation studies\cite{Hsu2014} and mechanical studies.\cite{Liu2014a}

More recently, the CVD growth technique has been used for direct growth of heterostructures.\cite{Gong2014} These \emph{epitaxially} grown heterostructures include TMDC-on-TMDC,\cite{Gong2014,Yu2015} \ce{MoS2}-on-\ce{hBN},\cite{Okada2014} TMDC-on-graphene,\cite{Ago2015} and \ce{MoS2}-on-\ce{SnS2}.\cite{Zhang2014c} Figure \ref{new_future}b shows an example of an epitaxially grown heterostructure. The outer triangular shape is single-layer \ce{MoS2} while the inner triangle is \ce{WS2}.\cite{Yu2015} These CVD grown heterostructures typically assemble as concentric triangles. 

Moreover, Gong \emph{et al}. have shown the possibility to grow lateral (in-plane) heterojunctions of different materials by changing the precursors in the reactor chamber.\cite{Gong2014} As shown by Yu \emph{et al}.\cite{Yu2015}, heterostructures built with transfer mechanisms show a reduced charge transfer compared to epitaxially grown ones; the charge transfer is equal after annealing the transferred heterostructures. To conclude, CVD techniques provide a reliable way to grow and fabricate heterostructures of layered semiconductors.

\subsection{Surface decoration} 
Surface decoration is a viable way to exploit the large surface-to-volume ratio to tune the properties of layered semiconductors. In the following paragraph we will discuss surface decoration with colloidal quantum dots. In the next section, which deals with light-matter interaction enhancement, we will also discuss decoration with gold nano structures for plasmonics.

\paragraph*{Hybrid \ce{MoS2}/Quantum-dot photodetectors.~~}
Very recently, Kufer \emph{et al.}\cite{Kufer2014} have realized hybrid photodetectors based on bilayer \ce{MoS2} and lead sulfide (\ce{PbS}) quantum dots. The final device was fabricated in two steps; first the bilayer \ce{MoS2} was patterned into an FET and, second, the \ce{PbS} quantum dots were deposited on top via a layer-by-layer and ligand exchange process. The ligand exchange step is crucial to improve the charge transport properties of the quantum dot film and the charge transfer to the \ce{MoS2} bilayer. The deposition of the QDs affects the \ids$-$\vg\ characteristics of the bilayer \ce{MoS2}, reducing the ON/OFF ratio and mobility. This is a strong indication that the QD deposition increases the charge carrier concentration in the \ce{MoS2} flake by charge transfer.

This hybrid photodetector shows high responsivity (\SI{6e5}{\A\per\W}), detection up to $\lambda$ \SI{= 1200}{\nm} (Figure \ref{new_future}c) and a time response in the order of \SI{350}{\ms}. Moreover, by tuning the \ce{MoS2} in the OFF state, the dark current can be suppressed, decreasing the NEP and increasing the detectivity of the detector. The low dark current and large detectivity are a major improvement with respect to a similar hybrid device based on single-layer graphene, where the large dark current of the graphene FETs reduces the detectivity.\cite{Konstantatos2012}

The photodetection mechanisms are a combination of photovoltaic and photo-gating. The electric field at the interface between the n-type \ce{MoS2} and the p-type \ce{PbS} quantum dot film separates the electron hole pairs that are generated by photon absorption in either of the two materials (photovoltaic). Next, the holes are trapped in the quantum dot film, increasing the photocarrier lifetime and yielding the large responsivity and long response time (photogating).

\subsection{Enhancing light-matter interaction}
Several techniques can be used to further increase light matter interaction in semiconductor devices. The topic of light matter interaction enhancement has been discussed in a recent perspective paper.\cite{Eda2013} In this Section, we only highlight three techniques that are relevant for 2D layered materials. 
 
\paragraph*{Plasmonic nanostructures.~~} To further boost the photoresponse, plasmonic surface decoration has been recently incorporated in detectors based on layered semiconductors.\cite{Britnell2013,Miao2015,Lin2013} Plasmonic structures enhance and focus the electric field of the optical excitation in a very small volume, effectively increasing the absorption cross section of the device. This enhancement has a very large effect on the device photoresponse, as can be seen in ref. \cite{Britnell2013} where the responsivity increases by an order of magnitude once the device is decorated with gold nanoparticles. 

\begin{figure*}[t]
\captionsetup{width=.9\textwidth}
\centering
\includegraphics[]{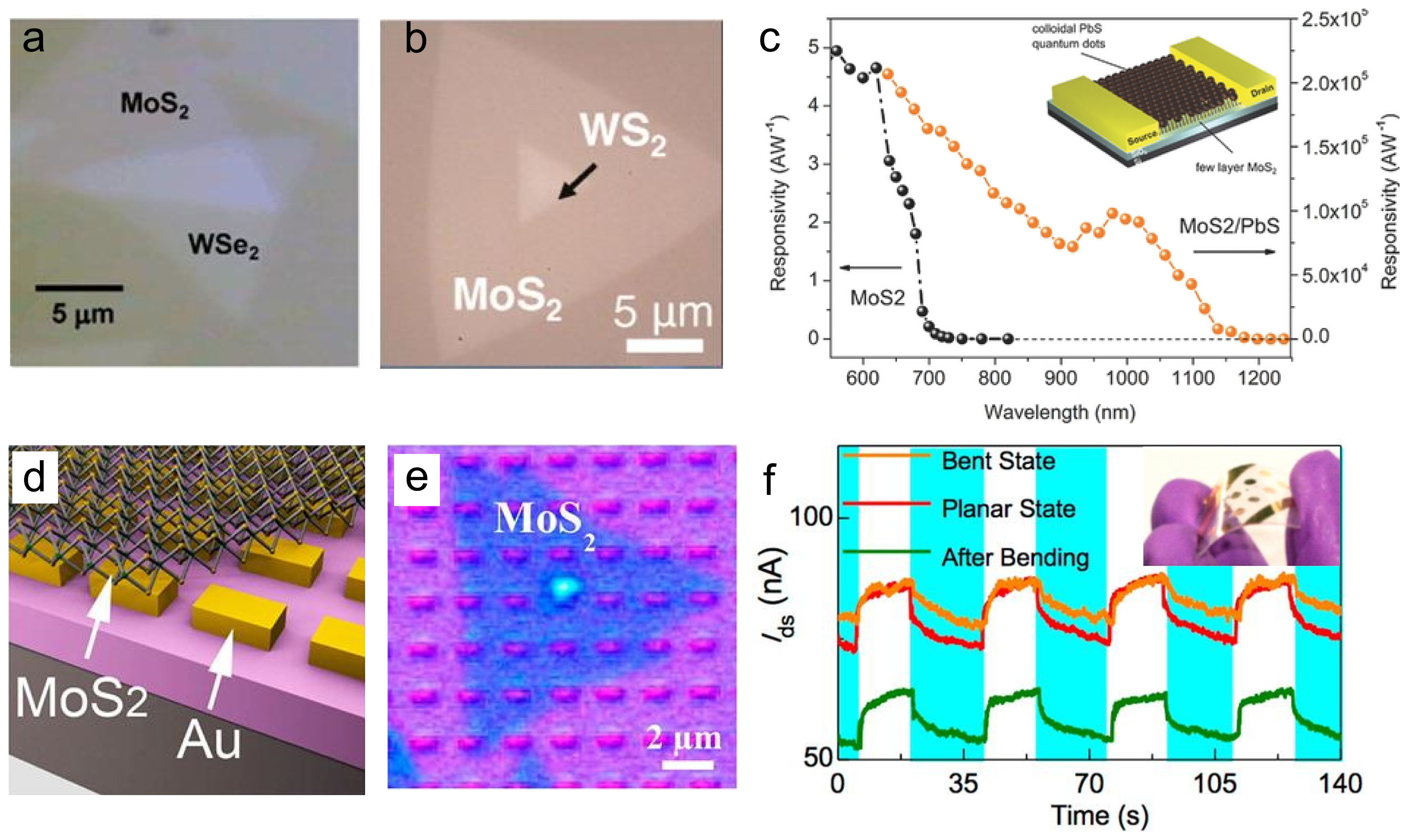}
\caption{(a) Optical micrograph of an artificial heterostructure built by vertically stacking single-layer \ce{MoS2} and \ce{WSe2} on sapphire. (b) Optical micrograph of epitaxially grown \ce{MoS2}/\ce{WS2} heterostructure. (c) Responsivity of a few-layer \ce{MoS2} phototransistor (black dots, left axis) and of the same transistor after decoration with \ce{PbS} quantum dots (orange dots, right axis). Schematic (d) and optical micrograph (e) of a single layer \ce{MoS2} deposited over a 2D array of gold nanoantennae. (f) Time trace of the source drain current of a flexible photodetector based on \ce{InSe}. The blue shading indicates that the optical excitation is OFF. The solid red line is the time-trace in the pristine state. The orange line is recorded while the device is bent. The green solid line is recorded after one bending cycle when the device is again flat. Panel a is adapted with permission from ref. \cite{Chiu2014}, copyright 2014, American Chemical Society. Panel b is adapted with permission from ref. \cite{Yu2015}, copyright 2015 American Chemical Society. Panel c is adapted with permission from ref. \cite{Kufer2014}, copyright 2014, John Wiley \& Sons. Panels d, e are adapted with permission from ref. \cite{Najmaei2014}, copyright 2014, American Chemical Society. Panel f and inset are adapted with permission from ref. \cite{Tamalampudi2014}, copyright 2014, American Chemical Society.}
\label{new_future}
\end{figure*}

Another example of plasmonic nanostructures is presented by an ordered array of metallic objects with dimension and pitch in the order of the excitation wavelength. Figure \ref{new_future}d and \ref{new_future}e show a schematic and an optical micrograph of such a structure consisting of a 2D square lattice of \num{100}-by-\num{1000} nm gold nanoantennas covered by a single-layer \ce{MoS2}.\cite{Najmaei2014} The asymmetry in the dimensions of the nanoantenna makes them sensitive to the polarization of the incoming excitation (or output emission). The pitch between the nanoantennae effects the center resonance wavelength of the plasmonic structure. Such structures can be relatively easily integrated onto photodetectors based on 2D semiconductors to increase their responsivity at a certain wavelength range. Conversely, they can also be applied to LEDs to increase their emission intensity.

\paragraph*{Photodetectors integrated onto waveguides.~~} Waveguides are planar structures that can guide an electric field in an analogue way to optical fibers, but confined in a plane. In silicon technology, they are being studied for on-chip optical interconnects at telecommunication wavelengths ($\lambda$ \SI{\sim1550}{\nm}) where \ce{Si} is transparent. At the boundaries of a waveguide, an evanescent field exists. By placing a thin flake of a layered semiconductor on top of a waveguide, this evanescent field can be absorbed by the material and give rise to electron-hole pairs useful for photodetection. Moreover, the interaction with the excitation electric field happens along the in-plane direction, drastically increasing the absorption cross section.

This concept has been used for graphene-based detectors, which achieved higher responsivities and large data-transfer rates.\cite{Gan2013,Wang2013a,Li2012b} To the best of our knowledge, the only other layered material that has been integrated on such a structure is black phosphorus in a research work by Youngblood \emph{et al}.\cite{Youngblood2015} that we have highlighted in a previous section. 

\paragraph*{Optical microcavities.~~} Other layered semiconductors (and especially the TMDCs) would not benefit from integration onto silicon waveguides due to their negligible absorption at telecommunication wavelengths. On the other hand, integration within optical microcavities in the visible greatly enhances light-matter interaction in such thin crystals. Very recently, a few groups have coupled \ce{MoS2},\cite{Liu2015} \ce{WSe2},\cite{Wu2014a,Wu2015} and GaSe \cite{Schwarz2014} to optical cavities to study the luminescence and demonstrate nanoscale optically-pumped lasing with very low threshold. Integration of a photodetector within an optical cavity would result in an increased responsivity at the cavity resonant wavelength, of interest for applications that require detection of a specific wavelength.

\subsection{Suspended devices} 

Studying devices based on suspended flakes of 2D semiconductors would allow the intrinsic properties of the semiconductor to be distinguished from spurious substrate-induced effects, such as optical interference, charge or phonon scattering. In a recent work, Klots \emph{et al}.\cite{Klots2014} perform photocurrent spectroscopy on suspended \ce{MoS2}, \ce{MoSe2} and \ce{WSe2} devices.  The photocurrent spectra they acquire allow the identification of the excitonic resonances that are responsible for photocurrent generation and extract their binding energy. To the best of our knowledge, this is the only study on the photocurrent of suspended devices. This is an interesting future direction for gaining a deeper fundamental understanding of the photoconduction processes. On the other hand, the difficulty arising in suspending the devices may be limiting the current experimental effort. We speculate that employing hBN as a gate dielectric would result in a suppression of impurity scattering and surface phonons (with respect to the common \ce{SiO2}) and, by carefully choosing the thickness, will also eliminate optical interference effects. Therefore, using h-BN as substrate may yield the same fundamental understanding as suspended devices.

\subsection{Flexible electronics} 
Flexible (nano)electronics is an emerging applied research field motivated by the need for bendable/foldable, transparent and functional electronic devices. Layered materials hold great promise to advance this field due to their atomic thickness, high crystalline quality (yielding high mobility), transparency (and yet strong light-matter interaction) and large breaking strength (allowing for large deformations). Another advantage of layered materials lays in the possibility of growing large-area of high crystalline quality material via chemical vapour deposition (CVD) techniques, as described earlier. A recent review highlights the impact of layered materials on the perspectives of flexible electronics.\cite{Akinwande2014}

Semiconducting layered materials could offer the basis for flexible transistors\cite{Zhu2015} or flexible tunneling transistors for digital computation\cite{Georgiou2013} and also the light-absorbing material for flexible photodetectors. \cite{Tamalampudi2014,Zhou2014,Amani2015,Sun2014a,Chen2015,Hu2013} 

Devices are typically built by transferring a flake of a layered semiconductor onto a flexible and transparent substrate, usually a polymer (a recent review on thin-film-based flexible electronics gives an overview of the available substrates\cite{Nathan2012}). Electrical contacts are already present on the substrate or they are patterned afterwards lithographically. An example of a bent device is shown in the inset of Figure \ref{new_future}f.\cite{Tamalampudi2014}. 

The performances of such detectors are less than an order of magnitude lower than their analogues on rigid \ce{SiO2}/\ce{Si} substrates. An interesting case is represented by a flexible photodetector based on few-layer \ce{GaS},\cite{Hu2013} as its performance is higher than the device built on \ce{SiO2}/\ce{Si}, stressing again the strong effect of the environment on 2D materials. These devices show some signs of degradation after bending cycles. 

Figure \ref{new_future}f shows measurements of the photoresponse of a \ce{In2Se3}-based flexible photodetector under modulated light excitation from ref. \cite{Tamalampudi2014}. The traces acquired in the pristine state and in the bent state are very similar, indicating that bending does not affect the photorespose. The green solid trace in Figure \ref{new_future}f is acquired once the bending stress is released and shows a very similar photocurrent and a lower OFF current. The similar photocurrent is again an indication that bending does not affect the photoresponse of \ce{In2Se3}. On the other hand, the reduced OFF current is likely an indication of an increased contact resistance due to the bending, most likely induced by degradation of the contact metals. 

Very recently, a flexible FET based on few-layer bP on metalized polyimide has shown unaltered electrical performances after 5000 bending cycles;\cite{Zhu2015} this is a promising indication that long-term stability to bending can be achieved. A further development would be to integrate flexible and transparent substrates with ionic polymer gating to combine the flexibility and transparency with an efficient field effect tunability.

To conclude, we would like to point out that not only photodetectors, but recently also LEDs have been realized on flexbile and transparent substrates.\cite{Withers2015} These impressive devices have been realized by employing only 2D materials: hBN as insulator and tunnel barrier, WS2 and MoS2 as semiconducting layers.

\section{Summary and outlook}

We have reviewed the current state-of-the-art in photodetection with layered semiconducting materials. Photodetectors based on TMDCs, especially \ce{MoS2}, show large responsivity coupled to slow response times, indicating that these materials can be suitable for sensitive applications in the visible and when time response is not important. \ce{MoS2} also possesses a large Seebeck coefficient, making it a promising material for thermal energy harvesting.
Novel materials, such as tri-chalcogenides (\ce{TiS3}) nanoribbons, III-VI and IV-VI compounds (e.g. \ce{GaTe} or \ce{SnS2}), show both large responsivities (larger than \ce{MoS2} photodetectors) and fast response times, making them promising candidates for fast and sensitive photodetection applications.
Few-layer black phosphorus and single-layer \ce{WSe2} show ambipolar transport and have been employed in the realization of electrical devices that can be controlled via local electrostatic gating. Black phosphorus has shown sizable and very fast photoresponse at telecommunication wavelengths with orders-of-magnitude larger responsivity of graphene and comparable response speed, making it an interesting material for optical communications and energy harvesting.

A future direction is represented by the possibility of stacking different 2D materials to form artificial heterostructures, allowing to tailor the resulting opto-electronic properties for a specific application. This has led to ultra-high responsive photodetectors (albeit very slow) based on graphene/\ce{MoS2} vertical stacks. Moreover, heterostructures of different TMDCS and/or graphene show great potential for solar energy harvesting in flexible and transparent solar cells. At this stage, a deeper understanding of the photocurrent and carrier recombination mechanisms is needed for optimization of device performance and large-area growth and transfer techniques will be fundamental for the realization of applications.

From an application perspective, the sizable variation in the figures-of-merit reported by different studies on the same material is an indication that device fabrication, contact metals and measurement environment play an important role in the photodetector performance. Furthermore, the fabrication process of these devices extensively relies on mechanical exfoliation. 
Both issues may be tackled by using CVD-grown materials, that provide large-area, uniform samples.
Integration with current CMOS technology is also an important challenge. Especially the work from Lopez-Sanchez \emph{et al.}\cite{Lopez-Sanchez2014,Lopez-Sanchez2014a} demonstrates that coupling \ce{MoS2} to silicon can be highly beneficial. Again, CVD growth may help in integrating \ce{MoS2} and other TMDCs in CMOS fabrication.

\section{Acknowledgement}
This work was supported by the Dutch organization for Fundamental Research on Matter (FOM). A.C-G. acknowledges financial support through the FP7-Marie Curie Project PIEF-GA-2011-300802 ('STRENGTHNANO').

\newpage
\footnotesize{
\bibliography{ref} 
\bibliographystyle{michele} 
}
\section{List of abbreviations}

\begin{table}[h!]

\centering

	\begin{tabular}{@{}rl@{}}\toprule
		Abbreviation & Description \\
		\midrule
		TMDCs & Transition metal dichalcodenides\\
		bP & Black phosphorus\\
		FET & Field-effect transistor\\
		$I_\mathrm{ds}$ & Source-drain current\\
		$I_\mathrm{ds}$ & Source-drain current\\
		$V_\mathrm{g}$ & Gate voltage\\
		PC & Photocondutance\\
		PG & Photogating\\
		PV & Photovoltaic\\
		PTE & Photothermoelectric\\
		SBs & Schottky Barriers\\
		SPCM & Scanning photocurrent microscopy\\
		$I_\mathrm{sc}$ & Short-circuit current\\
		$V_\mathrm{oc}$ & Open-circuit voltage\\
		$P_\mathrm{el}$ & Electrical power\\
		CVD & Chemical vapour deposition\\
		\bottomrule

	\end{tabular}
	\caption{List of abbreviations.}
\label{abbreviations}
\end{table}

\section{Appendix: Tables}

\begin{table*}

\centering

    \begin{tabular}{@{}lcccccccc@{}}\toprule
    \multirow{2}{*}{Material} & \multicolumn{4}{c}{Measurement conditions} & Responsivity & Rise time & \multirow{2}{*}{Spectral range} & \multirow{2}{*}{Reference} \\ 
    
    & \vds  (\si{\V}) & \vg  (\si{\V}) & $\lambda$  (\si{\nm}) & $P$  (\si{\mW\per\square\cm})& $R$ (\si{\respo}) & $\tau$ (\si{ms}) & & \\ \midrule
      
    {\color{white}$>$}1L \ce{MoS2} & \tnumb{1} & \tnumbb{50}& \num{532}&  \tnumbbb{8.0e4} & \tnumbbbb{7.5e0} & \tnumbbb{5e1} & Visible & \cite{Yin2012} \\
    
    {\color{white}$>$}1L \ce{MoS2} & \tnumb{8} & \tnumbb{-70}& \num{561} & \tnumbbb{2.4e-1}  & \tnumbbbb{8.8e5} & \tnumbbb{6e2} & Visible & \cite{Lopez-Sanchez2013} \\
    
    {\color{white}$>$}1L \ce{MoS2}\footnotesize{c} & \tnumb{1}& \tnumbb{41}& \num{532}& \tnumbbb{1.3e-1} & \tnumbbbb{2.2e6} & \tnumbbb{5e5} & - & \cite{Zhang2013a} \\
    
    {\color{white}$>$}1L \ce{MoS2} & \tnumb{1.5}& \tnumbb{0}& \num{514.5}& \tnumbbb{3.2e4} & \tnumbbbb{1.1e0} & \tnumbbb{1e4} & - & \cite{Perea-Lopez2014} \\
    
    $>$1L \ce{MoS2} & \tnumb{1}& \tnumbb{-2}& \num{633}& \tnumbbb{5.0e1} & \tnumbbbb{1.1e2} & \tnumbbb{1e3} & Visible - NIR& \cite{Choi2012} \\
    
    {\color{white}$>$}2L \ce{MoS2} & \tnumb{5}& \tnumbb{100}& WL\footnotesize{a} & $5$ \si{\nano\W} & \tnumbbbb{6.2e5} & \tnumbbb{2e4} & - & \cite{Furchi2014} \\   
        
    {\color{white}$>$}3L \ce{MoS2} & \tnumb{10}& \tnumbb{0}& \num{532}& \tnumbbb{2.0e3} & \tnumbbbb{5.7e2} & \tnumbbb{7.0e-2} & $\lambda$\ \SI{<700}{\nm} & \cite{Tsai2013} \\
                
    {\color{white}$>$}1L \ce{MoSe2} & \tnumb{10} & \tnumbb{0}& \num{532}&  \tnumbbb{1e2} & \tnumbbbb{1.3e1} & \tnumbbb{6e1} & - & \cite{Xia2014a} \\
    
    {\color{white}$>$}1L \ce{MoSe2} & \tnumb{1} & \tnumbb{0}& \num{650} & \tnumbbb{5.9e2}  & \tnumbbbb{2.6e-1} & \tnumbbb{2.5e1} & - & \cite{Chang2014} \\
    
  \:20L \ce{MoSe2} & \tnumb{20}& \tnumbb{0}& \num{532}& \tnumbbb{4.8e0} & \tnumbbbb{9.7e4} & \tnumbbb{3.0e1} & - & \cite{Abderrahmane2014} \\
      
       {\color{white}$>$}1L \ce{WS2}\footnotesize{b} & \tnumb{30}& \tnumbb{0}& \num{458}& \tnumbbb{2.0e0} & \tnumbbbb{2.1e-2} & \tnumbbb{5.3e0} & Visible & \cite{Perea-Lopez2013} \\

     {\color{white}$>$}1L \ce{WS2}\footnotesize{c} & \tnumb{30}& \tnumbb{0}& \num{633}& \tnumbbb{5.0e-2} & \tnumbbbb{1.3e4} & \tnumbbb{2.0e1} & - & \cite{Huo2014} \\  
     
     {\color{white}$>$}1L \ce{WSe2} & \tnumb{2}& \tnumbb{-60}& \num{650}& \tnumbbb{3.8e-1} & \tnumbbbb{1.8e5} & \tnumbbb{1.0e4} & - & \cite{Zhang2014} \\  
       
     {\color{white}$>$}1L \ce{WSe2} & \tnumb{0}& PN &  WL\footnotesize{a} & \tnumbbb{4.5e1} & \tnumbbbb{1.0e1} & - & - & \cite{Pospischil2014} \\ 
     
     {\color{white}$>$}1L \ce{WSe2} & \tnumb{0}& PN & \num{522}& \tnumbbb{1e-2} & \tnumbbbb{8.4e-1} & - & $\lambda$\ \SI{<820}{\nm} & \cite{Baugher2014} \\  
     
     {\color{white}$>$}1L \ce{WSe2} & \tnumb{0}& PN & \num{532}& \tnumbbb{8.4e-1} & \tnumbbbb{4e-1} & \tnumbbb{1.0e1} & $\lambda$\ \SI[separate-uncertainty]{<770 \pm 35}{\nm} & \cite{Groenendijk2014}  \\  
                   
    \bottomrule
    \multicolumn{9}{l}{\begin{minipage}[t]{0.69\columnwidth}%
 \footnotesize{a. WL = white light. b. Power in \si{\W}. c. Data reported for the device in vacuum.}
\end{minipage}}
    \end{tabular}
\caption{Figures-of-merit for \ce{MoS2}, \ce{MoSe2}, \ce{WS2} and \ce{WSe2} based photodetectors}
\label{materials_mos2_table}
\end{table*}

\begin{table*}

\centering

    \begin{tabular}{@{}lcccccccc@{}}\toprule
    \multirow{2}{*}{Material} & \multicolumn{4}{c}{Measurement conditions} & Responsivity & Rise time & \multirow{2}{*}{Spectral range} & \multirow{2}{*}{Reference} \\ 
    
    & \vds  (\si{\V}) & \vg  (\si{\V}) & $\lambda$  (\si{\nm}) & $P$  (\si{\mW\per\square\cm})& $R$ (\si{\respo}) & $\tau$ (\si{ms}) & & \\ \midrule
      
    \ce{TiS3} & \tnumb{1} & \tnumbb{-40}& \num{640}&  \tnumbbb{3e-1} & \tnumbbbb{2.9e6} & \tnumbbb{4} & $\lambda < 1100 \mathrm{nm}$ & \cite{Island2014} \\
    
    \ce{ZrS3} & \tnumb{5} & $-$ & \num{405}&  \tnumbbb{5e11} & \tnumbbbb{5e-2} & \tnumbbb{13e3} & $\lambda < 850 \mathrm{nm}$ & \cite{Tao2014} \\
    
    \ce{HfS3} & \tnumb{5} & $-$ & \num{405}&  \tnumbbb{1.2} & \tnumbbbb{1.1e2} & \tnumbbb{4e2} & $\lambda < 650 \mathrm{nm}$ & \cite{Xiong2014} \\

    \bottomrule
    
    \end{tabular}
\caption{Figures-of-merit for \ce{TiS3}, \ce{ZrS3} and \ce{HfS3} based photodetectors}
\label{materials_tric_table}
\end{table*}

\begin{table*}

\centering

 \begin{tabular}{@{}lcccccccc@{}}\toprule
    \multirow{2}{*}{Material} & \multicolumn{4}{c}{Measurement conditions} & Responsivity & Rise time & \multirow{2}{*}{Spectral range} & \multirow{2}{*}{Reference} \\ 
    & \vds  (\si{\V}) & \vg  (\si{\V}) & $\lambda$  (\si{\nm}) & $P$  (\si{\mW\per\square\cm})& $R$ (\si{\respo}) & $\tau$ (\si{ms}) & & \\ \midrule

	\ce{GaTe} & \tnumb{5}& \tnumbb{0}& \num{532}& \tnumbbb{3.0e-5} & \tnumbbbb{1e7} & \tnumbbb{6e0} & Visible & \cite{Liu2013} \\
    
	\ce{GaSe} & \tnumb{5}& \tnumbb{0}& \num{254}& \tnumbbb{1.0e0} & \tnumbbbb{2.8e3} & \tnumbbb{3e2} & UV - Visible & \cite{Hu2012} \\
    
	\ce{GaS} & \tnumb{2}& \tnumbb{0}& \num{254}& \tnumbbb{2.6e-2} & \tnumbbbb{4.2e3} & \tnumbbb{3e1} & UV - Visible & \cite{Hu2013} \\
    
	\ce{In2Se3} & \tnumb{5}& \tnumbb{0}& \num{300}& \tnumbbb{2.1e-1} & \tnumbbbb{3.9e5} & \tnumbbb{1.8e1} & UV - NIR & \cite{Jacobs-Gedrim2013} \\
    
	\ce{InSe} & \tnumb{10}& \tnumbb{80}& \num{633}& \tnumbbb{3.5e2} & \tnumbbbb{1.6e5} & \tnumbbb{4.0e3} & Visible - NIR & \cite{Tamalampudi2014} \\
        
	\ce{InSe} & \tnumb{3}& \tnumbb{0}& \num{532}& \tnumbbb{2.5e2} & \tnumbbbb{3.5e1} & \tnumbbb{5.0e-1} & Visible - NIR & \cite{Lei2014} \\    
    
	\ce{SnS2} & \tnumb{3}& \tnumbb{0}& \num{457}& \tnumbbb{2e3} & \tnumbbbb{9} & \tnumbbb{5e-3} & Visible - NIR & \cite{Su2014} \\    
        
    \bottomrule

    \end{tabular}
\caption{Figures-of-merit for photodetectors based on III-VI compounds}
\label{novel_mat}
\end{table*}

\begin{table*}
\centering

    \begin{tabular}{@{}lcccccccc@{}}\toprule
    \multirow{2}{*}{Thickness (\si{\nm})} & \multicolumn{4}{c}{Measurement conditions} & Responsivity & Rise time & \multirow{2}{*}{Spectral range} & \multirow{2}{*}{Reference} \\ 
    
    & \vds  (\si{\V}) & \vg  (\si{\V}) & $\lambda$  (\si{\nm}) & $P$  (\si{\mW\per\square\cm})& $R$ (\si{\respo}) & $\tau$ (\si{ms}) & & \\ \midrule
      
     \tnumb{8} & \tnumb{0.2} & \tnumbb{0} & \num{640}& \tnumbbb{1.6e1} & \tnumbbbb{4.8e0} & \tnumbbb{1e0} & Visible - NIR & \cite{Buscema2014}\\
     
     \tnumb{6} & \tnumb{0} & PN & \num{532}& \tnumbbb{1.9e3} & \tnumbbbb{5e-1} & \tnumbbb{1.5e0} & Visible - NIR & \cite{Buscema2014b} \\
     
     \tnumb{6} & \tnumb{-0.5} & PN & \num{532}& \tnumbbb{1.9e3} & \tnumbbbb{2.8e1} & - & Visible - IR & \cite{Buscema2014b} \\
    
    \tnumb{120} & \tnumb{0} & \tnumbb{0} & \num{1550}& \tnumbbb{3e6} & \tnumbbbb{5.0e0} & - & Visible - IR & \cite{Engel2014a, Low2014a} \\
     
      \tnumb{120} & \tnumb{0} & \tnumbb{0} & \num{532}& \tnumbbb{1e7} & \tnumbbbb{2.0e1} & - & Visible - NIR & \cite{Engel2014a, Low2014a} \\
    
    \tnumb{11.5} & \tnumb{0.4} & \tnumbb{-8} & \num{1550}& \tnumbbb{1.91e0} & \tnumbbbb{1.3e2} & a & Visible - NIR & \cite{Youngblood2015}b \\

    \bottomrule
    \multicolumn{9}{l}{\begin{minipage}[t]{0.69\columnwidth}%
 \footnotesize{a. $f_\mathrm{3dB}$\ \SI{=3}{\GHz} b. Power in \si{\milli\W}. }
\end{minipage}}
    \end{tabular}
\caption{Figures-of-merit for few-layer bP based photodetectors}
\label{materials_bP_table}
\end{table*}

\end{document}